\documentclass[review,english]{elsarticle}

\journal{Journal of Magnetic Resonance}
\biboptions{sort&compress}
\usepackage{numcompress}
\usepackage[T1]{fontenc}
\usepackage[latin1]{inputenc}
\usepackage[percent]{overpic}
\usepackage{pict2e}
\usepackage{graphicx}
\usepackage{booktabs}
\usepackage{threeparttable}
\usepackage{float}
\usepackage{csquotes}

\usepackage{algorithm}
\usepackage{algcompatible}
\usepackage{amsmath,amsfonts,amssymb}
\usepackage{graphicx}
\usepackage{diagbox}

\usepackage[usenames,dvipsnames,table]{xcolor}

\usepackage{soul}

\mathchardef\mhyphen="2D

\usepackage{babel}
\makeatother
\usepackage{url}
\begin{document}

\begin{frontmatter}

\title{4D-image reconstruction directly from limited-angular-range data in continuous-wave electron paramagnetic resonance imaging}

\author[a]{Zheng Zhang}
\author[b]{Boris Epel}
\author[a]{Buxin Chen}
\author[a]{Dan Xia}
\author[a]{Emil Y. Sidky}
\author[c]{Zhiwei Qiao}
\author[b]{Howard Halpern}
\author[a,b]{Xiaochuan Pan\fnref{myfootnote}}
\address[a]{Department of Radiology, The University of Chicago, Chicago, IL, USA}
\address[b]{Department of Radiation \& Cellular Oncology, The University of Chicago, Chicago, IL, USA}
\address[c]{School of Computer and Information Technology, Shanxi University, Taiyuan, Shanxi, China}
\fntext[myfootnote]{Corresponding author. Email: xpan@uchicago.edu.}


\begin{abstract}
{\bf Objective}: We investigate and develop optimization-based algorithms for accurate reconstruction of four-dimensional (4D)-spectral-spatial (SS) images directly from data collected over limited angular ranges (LARs) in continuous-wave (CW) electron paramagnetic resonance imaging (EPRI). 
{\bf Methods}: Basing on a discrete-to-discrete data model devised in CW EPRI employing the Zeeman-modulation (ZM) scheme for data acquisition, we first formulate the image reconstruction problem as a convex, constrained optimization program that includes a data fidelity term and also constraints on the individual directional total variations (DTVs) of the 4D-SS image. Subsequently, we develop a primal-dual-based DTV algorithm, simply referred to as the DTV algorithm,
to solve the constrained optimization program for achieving image reconstruction from data collected in LAR scans in CW-ZM EPRI. 
{\bf Results}: We evaluate the DTV algorithm in simulated- and real-data studies for a variety of LAR scans of interest in CW-ZM EPRI, and visual and quantitative results of the  studies reveal that 4D-SS images can be reconstructed directly from LAR data, which are visually and quantitatively comparable to those obtained from data acquired in the standard, full-angular-range (FAR) scan in CW-ZM EPRI. 
{\bf Conclusion}: An optimization-based DTV algorithm is developed for accurately reconstructing 4D-SS images directly from LAR data in CW-ZM EPRI.  Future work includes the development and application of the optimization-based DTV algorithm for reconstructions of 4D-SS images from FAR and LAR data acquired in CW EPRI employing schemes other than the ZM scheme.
{\bf Significance}: The DTV algorithm developed may be exploited potentially for enabling and optimizing CW EPRI with minimized imaging time and artifacts by acquiring data in LAR scans.
\end{abstract}


\end{frontmatter}


\section{Introduction}\label{sec:introduction}
Electron paramagnetic resonance imaging (EPRI) has found  applications to biological and preclinical imaging \cite{berliner1985magnetic,kuppusamy1994three,elas2003quantitative,eaton2004biomedical,matsumoto2006electron, elas2008electron,epel2011comparison,biller2016rapid,epel2017vivo,eaton2018epr}. In EPRI, one seeks to determine quantitatively the four-dimensional (4D)-spectral-spatial (SS) distribution of paramagnetic electron-spins of contrast materials, referred to as the 4D-SS image, from which physical quantities of biological or clinical interest such as oxygen concentration within the imaged subject may be determined \cite{maltempo1987spectral, kuppusamy1995mapping,mader1997non, kuppusamy1998vivo,williams2005epr, pan2007targeted}. Algorithms have been developed for reconstructing 4D-SS images from data acquired in the full-angular-range (FAR) scan in CW EPRI \cite{tseitlin2007comparison,tseitlin2008regularized,tseitlin2014new,komarov2017fast}. It is of theoretical and practical interest in reconstruction of 4D-SS images from data acquired in limited-angular-range (LAR) scans in CW EPRI, and we thus investigate and develop  algorithms for accurate reconstruction of 4D-SS images directly from data collected in LAR scans in CW EPRI. 

Among various scanning schemes, the Zeeman modulation (ZM)  scheme \cite{poole1996electron,elas2003quantitative,subramanian2004radio,williams2005epr} has been used to acquire data in CW EPRI.  The filtered-backprojection (FBP) algorithm is used currently for reconstructing the partial derivative (PD) of a 4D-SS image with respect to its spectral dimension \cite{williams2005epr,som2007parametric} from data collected in the FAR scan in CW-ZM EPRI; and carrying numerically out an accumulative integration of the 4D-PD-SS image over the spectral dimension, one subsequently obtains a 4D-SS image. In the work, we focus on the development and investigation of an optimization-based algorithm for accurate reconstruction of 4D-SS images directly from data acquired in FAR and LAR scans in CW-ZM EPRI. The approach and algorithm developed can, however, readily be extended to reconstruct 4D-SS images from data collected in FAR and LAR scans in CW EPRI that employs other schemes than the ZM scheme for data acquisition. We refer to data collected, respectively, in FAR and LAR scans simply as {\it FAR data} and {\it LAR data} hereinafter.

One of the motivations of our work is based upon the observation of the physical constraint on the gradient magnetic field, which is used for achieving spatial encoding in CW EPRI. The gradient strength applied specifies the tangent of the angle between the directions of the main magnetic field and a sampling hyperplane in the 4D-SS space. When the angle is in the vicinity of $\pm 90^\circ$, its tangent approaches infinity, and thus no physical measurement at $\pm 90^\circ$ can be made because it requires an infinitely large gradient strength, which is physically impossible. Therefore, reconstruction of a 4D-SS image in practical CW EPRI can thus be an inherently LAR problem. In particular, in practical CW-ZM EPRI \cite{williams2005epr}, noise and other physical perturbation in data acquired in the vicinity of $\pm 90^\circ$ may be amplified in reconstruction of a 4D-SS image with existing algorithms such as the FBP algorithm \cite{ahn2007simulation,som2007parametric}, yielding a 4D-SS image of sub-optimal quality.


In CW-ZM EPRI, the standard data model that relates a 4D-PD-SS image to data can be interpreted as the 4D-Radon transform of the 4D-PD-SS image \cite{williams2005epr}, as discussed in
\ref{sec:cc-model}, and existing algorithms such as the FBP algorithm are used thus for reconstructing the 4D-PD-SS image from FAR data. There exists, however, strong interests in obtaining 4D-SS images directly from FAR and LAR data in CW-ZM EPRI. 
It remains unclear, to the best of our knowledge, that algorithms exist for reconstruction of a 4D-SS image directly from FAR or LAR data in CW-ZM EPRI. We investigate and develop such algorithms in the work.
 
We first formulate the image-reconstruction problem as a convex, constrained optimization program, which includes a data-fidelity term and also constraints on the directional total variations (DTVs) of the 4D-SS image. The choice of the DTV constraints is motivated by a recent work that demonstrates the potential utility of DTV constraints for accurate image reconstruction from LAR data in computed tomography (CT) \cite{zhang2021directional,chen2021dual}. We then develop a  primal-dual-based DTV algorithm, which is referred to simply as the DTV algorithm,
to reconstruct a 4D-SS image directly from FAR and LAR data by solving the convex, constrained optimization problem formulated in CW-ZM EPRI.

We perform simulated- and real-data studies to verify and evaluate the effectiveness and stability of the DTV algorithm developed for reconstruction of 4D-SS images directly from FAR and LAR data acquired in CW-ZM EPRI. Results of the studies reveal that the DTV algorithm developed can reconstruct accurately 4D-SS images directly from FAR data, and also from LAR data with visual appearance and numerical quantification comparable to their counterparts obtained from FAR data in CW-ZM EPRI. 

Following the introduction in Sec. \ref{sec:introduction}, we discuss in Sec. \ref{sec:method} a data model in practical CW-ZM EPRI, a formulation of the reconstruction problem as an optimization program, the development of the DTV algorithm solving the program, and strategies for verifying and evaluating the DTV algorithm. Results of simulated- and real-data studies, including 4D-SS images reconstructed by use of the DTV algorithm directly from FAR and LAR data, are presented in Sec. \ref{sec:results}, followed by discussion and conclusion in Secs. \ref{sec:discussion} and \ref{sec:conclusion}. 

\section{Materials and methods}\label{sec:method}

\begin{figure}
\centering
\includegraphics[angle=0,trim=20 80 20 0, clip,origin=c,width=0.9\textwidth]{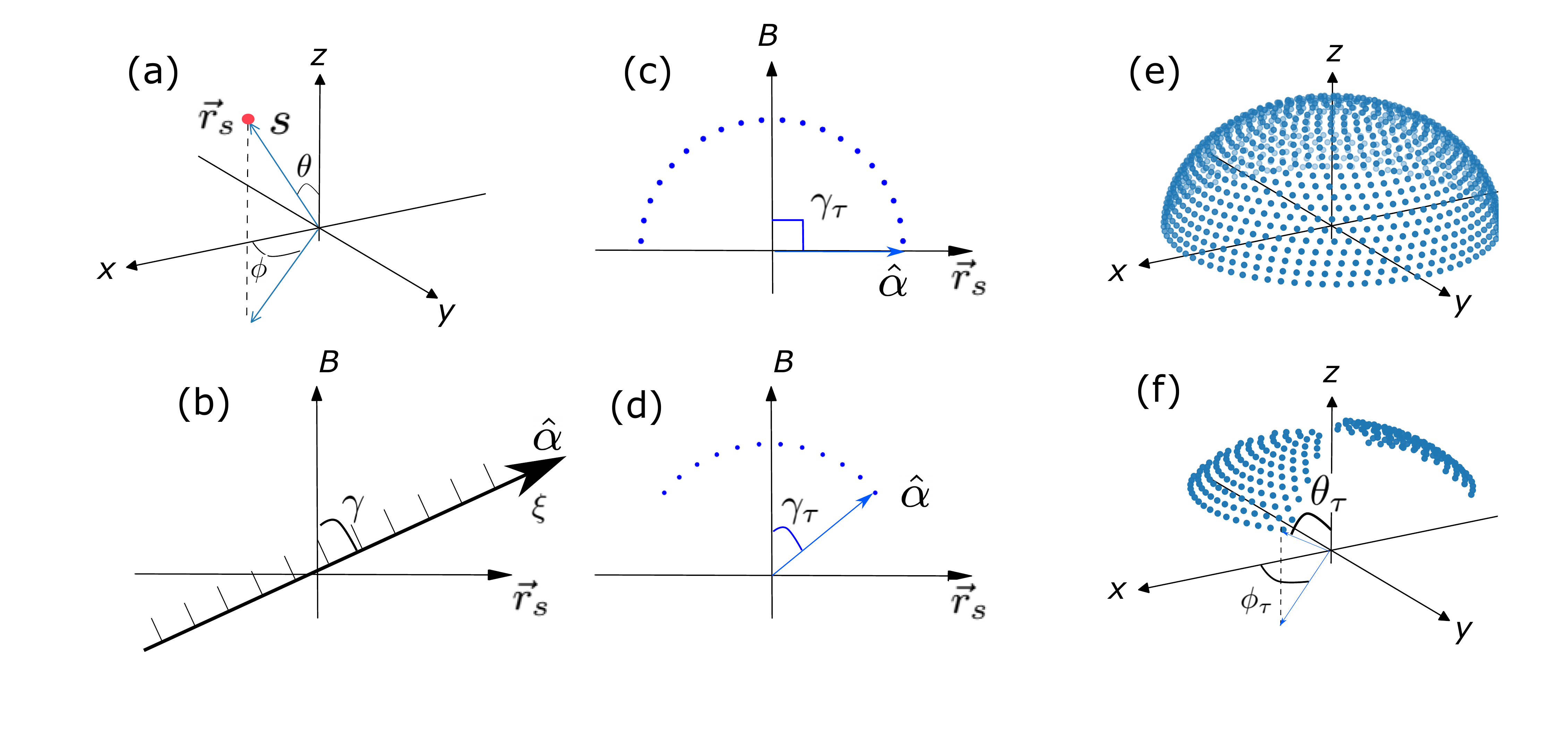}
\caption{{(a) a 3D-spatial coordinate system in which the orientation of vector $\vec{r}_s$, pointing to a spatial point $s$, is specified by angles $\theta$ and $\phi$; (b) an effective 2D space formed by two orthogonal dimensions  $\vec{r}_s$ and $B$ (i.e., representing 4D space $\{x, y, z, B\})$ in which 4D unit vector $\hat{\alpha}$ has angle $\gamma$ relative to the $B$-axis; (c) and (d): uniform sampling grids over $\gamma$ in FAR and LAR scans with $\gamma_\tau=90^\circ$ and $60^\circ$; and ESA-sampling grids over the entire hemisphere (e)  specified by $\theta_\tau=90^\circ$ and $\phi_\tau=90^\circ$, and over the partial hemisphere (f) specified by $\theta_\tau=60^\circ$ and $\phi_\tau=45^\circ$,} in the 3D-spatial space $\{x, y, z\}$ for the FAR scan and a LAR scan described in Sec. \ref{sec:data-acquisition}.}
\label{fig:config}
\end{figure}

Let three-dimensional (3D) vector $\vec{r}_s=(r_s {\rm cos}\phi \,{\rm sin}\theta, r_s {\rm sin}\phi \,{\rm sin}\theta, r_s {\rm cos}\theta)^{\top}$ depicts a point in the 3D-spatial-coordinate system, where $r_s$ denotes the magnitude of $\vec{r}_s$, ``${\top}$'' the transpose, $\theta$ the angle between $z$-axis and $\vec{r}_s$, and $\phi$ the angle between the $x$-axis and the projection of $\vec{r}_s$ onto the $x$-$y$ plane, as shown in Fig. \ref{fig:config}a.
Also, letting $B$ denote the magnetic field applied, we introduce 4D vector $\vec{r}=(x, y, z, B)^{\top}$  to denote a point in 4D-image space. We use a 4D unit-vector $\hat{\alpha}=({\rm cos}\phi \,{\rm sin}\theta\,{\rm sin}\gamma, {\rm sin}\phi \,{\rm sin}\theta\,{\rm sin}\gamma,$ ${\rm cos}\theta\,{\rm sin}\gamma, {\rm cos}\gamma)^{\top}$ to define the orientation of a hyperplane in the 4D space, where $\gamma$ depicts the angle between the $B$-axis and $\hat{\alpha}$, as shown in Fig. \ref{fig:config}b. We use $f(\vec{r})$ to denote a continuous 4D-SS image as 4D variable $\vec{r}$ can continuously vary. Also, we use $g(\xi,\hat{\alpha})$ to denote a continuous 4D-data function in a 4D space formed by scalar variable $\xi$ and 4D-unit-vector $\hat{\alpha}$, where $\xi$ depicts the distance of a hyperplane of orientation $\hat{\alpha}$ to the origin in the 4D-image space, and $\xi$ and $\hat{\alpha}$ can vary continuously.
As shown in Sec. \ref{sec:cc-model}, for CW-ZM EPRI, $g(\xi,\hat{\alpha})$ is related to the 4D-Radon transform, i.e., the integration, of $\frac{\partial f(\vec{r})}{\partial B}$ computed over the hyperplane, where $\frac{\partial f(\vec{r})}{\partial B}$ indicates the partial derivative of continuous 4D-SS image $f(\vec{r})$ with respect to $B$. As discussed in Sec. \ref{sec:cc-model}, a continuous-to-continuous (CC)-data model shown in Eq. \eqref{eq:cc-model} can be established for CW-ZM EPRI \cite{williams2005epr}.

\subsection{Discrete-to-discrete (DD)-data model in CW-ZM EPRI}\label{sec:dd-data-form}

In realistic CW-ZM EPRI, both image and data are presented on discrete arrays, and one thus discretizes the variables of $f(\vec{r})$ and $g(\xi,\hat{\alpha})$ onto discrete 4D-image and 4D-data arrays. Specifically, in the work, variable $\vec{r}$ of $f(\vec{r})$ is discretized onto a discrete array of $I$ identical 4D voxels, where $I=I_x\times I_y \times I_z\times I_B$, and $I_x$, $I_y$, $I_z$, and $I_B$ depict, respectively, the array sizes along dimensions $x$, $y$, $z$, and $B$. Each of the 4D voxels is of identical size $\Delta_x\times \Delta_y\times \Delta_z\times \Delta_B$, respectively, filling completely in the 4D support of $L \times L \times L \times B_w$, i.e., intervals $\Delta_x={L}/I_x$,  $\Delta_y={L}/I_y$, $\Delta_z={L}/I_z$, and $\Delta_B={B}_w/I_B$. We use vector $\mathbf{f}$ of size $I$ to denote the 4D-SS image represented on the 4D-image array in a concatenated form in the order of $x$, $y$, $z$, and $B$, and entry $i$ of vector $\mathbf{f}$ denotes the value of the 4D-SS image in voxel $i$, where $i=1, 2, ..., I$.  For simplicity, we refer  to {\it discrete 4D-SS image $\mathbf{f}$ as an SS image} hereinafter. 

Similarly, variables $\xi$ and $\hat{\alpha}$ of 4D-data function $g(\xi,\hat{\alpha})$ can be discretized onto a discrete array of $J$ grids, where $J=J_\alpha \times J_\xi$, $J_\alpha$ denotes the total number of  discrete  samples each of which is specified by $\hat{\alpha}$ on the hypersphere in the 4D-image space, and let $J_\xi$ depict, for each $\hat{\alpha}$, the total number of samples over $\xi\in [-\xi_{\tau}, \xi_{\tau}]$, where $2 \xi_\tau$ depicts the maximum support of the data function in the $\xi$-dimension. Also, using $\gamma_{\tau}\le 90^\circ$,  $\theta_{\tau}\le 90^\circ$, and  $\phi_{\tau}\le 90^\circ$ to denote the maximum angles in a scan, we have $\gamma\in [-\gamma_{\tau}, \gamma_{\tau}]$, $\theta\in [-\theta_{\tau}, \theta_{\tau}]$, and $\phi\in [-\phi_{\tau},\phi_{\tau}]$. 
Assuming that sample grids are obtained by varying $\hat{\alpha}$ in the order of varying $\phi$, $\theta$, and  $\gamma$ and then by varying $\xi$, we can then use vectors $\mathbf{g}$ and $\mathbf{g}^{[\mathcal{M}]}$ of size $J$ to denote discrete model data and measured data, respectively, in a concatenated form in the order of $\phi$,  $\theta$, $\gamma$, and $\xi$. For simplicity, we refer henceforward to $\mathbf{g}$ and $\mathbf{g}^{[\mathcal{M}]}$,  as the {\it model data} and {\it measured data}, respectively.

Basing upon the CC-data model in Eq. \eqref{eq:cc-model} and discrete image and data arrays devised above,  we devise a discrete-to-discrete (DD)-data model, which relates $\mathbf{f}$ and $\mathbf{g}$ defined above, as
\begin{eqnarray}\label{eq:dd-model-A}
\mathbf{g} = \mathcal{H} \mathbf{f}
\quad\quad {\rm with} \quad\quad \mathcal{H} = \mathcal{C}\mathcal{R}\mathcal{D}_B,
\end{eqnarray}
where system matrix $\mathcal{H}$ is of size $J \times I$, matrix $\mathcal{R}$ of size $J \times I$ denotes the 4D-Radon transform in a discrete form in which element $r_{ji}$ is chosen to be the intersection hyperarea of hyperplane $j$ with voxel $i$ in the 4D-spectral-spatial space; $\mathcal{C}$ a diagonal  matrix of size $J \times J$ in which diagonal element 
$c_{j j}={\rm cos} \gamma_j$; $\gamma_j$
the angle between the $B$-axis and vector $\hat{\alpha}_j$; $\hat{\alpha}_j$ the orientation of a hyperplane at the $j$th projection; and matrix $\mathcal{D}_B$ of size $I \times I$ represents two-point difference along spectral dimension $B$. We note that 
$\vert \gamma_j\vert \le \gamma_{\tau}$ and that $\gamma_{\tau}<90^\circ$ because it is physically impossible to collect data at $\gamma=90^\circ$ in CW EPRI. 

\subsection{Equal-solid-angle (ESA)-sampling scheme}\label{sec:data-acquisition}
The explicit form of the DD-data model, i.e., $\mathcal{H}$, depends upon not only ranges $\xi_\tau$, $\gamma_\tau$, $\theta_\tau$, and $\phi_\tau$ but also the specific sampling schemes (i.e., the grids) over the ranges. In Fig. \ref{fig:config}a, we use 3D vector $\vec{r}_s$ to denote a point in 3D-spatial space $\{x, y, z\}$. For a FAR or LAR scan in CW-ZM EPRI, system matrix $\mathcal{H}$ in the DD-data model in Eq. \eqref{eq:dd-model-A} depends upon how the 4D-data space is sampled. In the work, we divide its $\xi$-axis into a total of $J_\xi$ samples uniformly distributed over $\xi\in [-\xi_{\tau}, \xi_{\tau}]$ separated by constant interval $\Delta_{\xi}={2 \xi_{\tau}}/J_\xi$, as shown in Fig. \ref{fig:config}b; and we then consider below the equal-solid-angle (ESA)-sampling scheme that has been used for data acquisition in CW-ZM EPRI \cite{epel2011comparison}.

In an ESA-sampling scheme, $J_\gamma$ samples are evenly distributed over $[-\gamma_{\tau}, \gamma_{\tau}]$ with constant angular interval $\Delta_\gamma$, as shown in Figs. \ref{fig:config}c and  \ref{fig:config}d, whereas $J_{\theta}$ samples evenly distributed over $[-\theta_{\tau}, \theta_{\tau}]$ with constant angular interval $\Delta_\theta$ and  $J_{\phi}(\theta_n)$ samples evenly distributed
over $\phi$ with  angular interval $\Delta_\phi(\theta_n)={\rm min}(|\frac{\Delta_\gamma}{{\rm sin} (\theta_n)}|,\pi)$, where function ${\rm min}(\cdot)$ returns the smallest value, yielding the respective ESA-sampling grids shown. 
For example, for the FAR scan, $\gamma_{\tau}=90^\circ$, $\theta_{\tau}=90^\circ$, and $\phi_{\tau}=90^\circ$, and its ESA sampling grids are displayed in Figs. \ref{fig:config}c and \ref{fig:config}e, whereas for a LAR scan specified by $\gamma_{\tau}=60^\circ$, $\theta_{\tau}=60^\circ$, and $\phi_{\tau}=45^\circ$, its ESA sampling grids are depicted in Figs. \ref{fig:config}d and \ref{fig:config}f.

\subsection{Constrained optimization program}


Image reconstruction in CW-ZM EPRI is tantamount to invert $\mathcal{H}$ for obtaining SS image  $\mathbf{f}$ from measured data $\mathbf{g}^{[\mathcal{M}]}$ in a scan.
In an attempt to invert the DD-data model in Eq. \eqref{eq:dd-model-A}, we formulate its inversion, i.e., the reconstruction problem of $\mathbf{f}$, as a convex, constrained optimization program  \cite{zhang2021directional,chen2021dual,zhang2022image} given by
\begin{eqnarray}\label{eq:optA}
\mathbf{f}^{\star}\! &=&\! \underset{\mathbf{f}}{\mathsf{argmin}} \!\left \{\frac{1}{2} \!\parallel \mathcal{C}\mathcal{R}\mathcal{D}_B {\mathbf{f}} - \mathbf{g}^{[\mathcal{M}]} \parallel_2^2 \right \} \\ \nonumber 
& & {\rm s.t.} \quad ||\mathcal{D}_x 
{\mathbf{f}}||_1  \!\le  \!t_x, \,\, ||\mathcal{D}_y {\mathbf{f}}||_1  \!\le  \! t_y, \,\, ||\mathcal{D}_z {\mathbf{f}}||_1  \!\le  \!t_z, \,\, ||\mathcal{D}_B {\mathbf{f}}||_1 \! \le \! t_B,  \,\, f_i \! \ge \! 0,
\end{eqnarray}
where matrices $\mathcal{D}_x$, $\mathcal{D}_y$, $\mathcal{D}_z$, and $\mathcal{D}_B$ of sizes $I \times I$ denote two-point differences along spatial dimensions $x$, $y$, and $z$ and spectral dimension $B$, respectively; 
$||\cdot||_1$ depicts the $\ell_1$-norm of the input vector; $||\mathcal{D}_x \mathbf{f}||_1$, $||\mathcal{D}_y \mathbf{f}||_1$, $||\mathcal{D}_z \mathbf{f}||_1$, and $||\mathcal{D}_B \mathbf{f}||_1$ are  referred to, respectively, as the DTVs of SS image $\mathbf{f}$ along spatial dimensions $x$, $y$, and $z$ and spectral dimension $B$; and constraint parameters $t_x$, $t_y$, $t_z$, and $t_B$ are the upper bounds on the respective DTVs of SS image $\mathbf{f}$. Because a physical SS image is non-negative, an additional constraint, i.e., $f_i \ge 0$, is included in the optimization program in Eq. \eqref{eq:optA}.

\subsection{DTV algorithm for reconstruction of SS image $\mathbf{f}$}\label{sec:DTV_algorithm}
Basing upon the primal-dual approach to solving convex optimization problems \cite{c-p:2011,sidky2012convex,zhang2021directional}, we derive a primal-dual instance algorithm, referred to as the DTV algorithm, to reconstruct SS image $\mathbf{f}$ through solving the optimization program in Eq. \eqref{eq:optA}, and show the pseudo-codes of the DTV algorithm in Algorithm \ref{alg:1}.

\begin{algorithm}\leavevmode
\caption{Pseudocode of the DTV algorithm reconstructing SS image $\mathbf{f}$ through solving Eq. \eqref{eq:optA}}\label{alg:1}
\begin{algorithmic}[1]
\STATEx INPUT: $g^{[\mathcal{M}]}$, ${\mathcal{H}}$, ${t}_x$, ${t}_y$, ${t}_z$, ${t}_B$, and $\rho\approx 10^{-2}$
\STATE $\nu_1 \leftarrow \sqrt{e_L}/||\mathcal{D}_x||_2$, $\nu_2 \leftarrow \sqrt{e_L}/||\mathcal{D}_y||_2$, $\nu_3 \leftarrow \sqrt{e_L}/||\mathcal{D}_z||_2$, $\nu_4 \leftarrow \sqrt{e_L}/||\mathcal{D}_B||_2$,  $\mu \leftarrow \sqrt{e_L}||_2/||\mathcal{I}||_2$, $\Sigma$, $\mathcal{T}$
\STATE $n \leftarrow 0$
\STATE INITIALIZE: ${\mathbf{f}}^{(0)}$, $\mathbf{w}^{(0)}$, $\mathbf{p}^{(0)}$, $\mathbf{q}^{(0)}$, $\mathbf{r}^{(0)}$, $\mathbf{b}^{(0)}$, and $\mathbf{t}^{(0)}$ to zero
\STATE $\bar{\mathbf{f}}^{(0)} \leftarrow \mathbf{f}^{(0)}$
\REPEAT
\STATE  $\mathbf{w}^{(n+1)} = (\mathbf{w}^{(n)} + \Sigma(\mathcal{H}\bar{\mathbf{f}}^{(n)} - \mathbf{g}^{[\mathcal{M}]}))/(\mathcal{I}+\Sigma)$
\STATE $\bar{\mathbf{p}}^{(n)} = \mathbf{p}^{(n)} + \Sigma \nu_1 \mathcal{D}_x \bar{\mathbf{f}}^{(n)}$
\STATEx \hspace{0.25cm} $\bar{\mathbf{q}}^{(n)} = \mathbf{q}^{(n)} + \Sigma \nu_2 \mathcal{D}_y \bar{\mathbf{f}}^{(n)}$
\STATEx \hspace{0.25cm} $\bar{\mathbf{r}}^{(n)} \hspace{0.1cm} = \mathbf{r}^{(n)} + \Sigma \nu_3 \mathcal{D}_z \bar{\mathbf{f}}^{(n)}$
\STATEx \hspace{0.25cm} $\bar{\mathbf{b}}^{(n)} = \mathbf{b}^{(n)} + \Sigma \nu_4 \mathcal{D}_B \bar{\mathbf{f}}^{(n)}$
\STATE $\mathbf{p}^{(n+1)} = \bar{\mathbf{p}}^{(n)} - \Sigma \frac{\bar{\mathbf{p}}^{(n)}}{|\bar{\mathbf{p}}^{(n)}|}\ell_1 {\rm ball}_{\nu_1 {t}_x} (\frac{\bar{\mathbf{p}}^{(n)}}{\Sigma})$
\STATEx \hspace{0.25cm} $\mathbf{q}^{(n+1)} = \bar{\mathbf{q}}^{(n)} - \Sigma \frac{\bar{\mathbf{q}}^{(n)}}{|\bar{\mathbf{q}}^{(n)}|}\ell_1 {\rm ball}_{\nu_2 {t}_y} (\frac{\bar{\mathbf{q}}^{(n)}}{\Sigma})$
\STATEx \hspace{0.25cm} $\mathbf{r}^{(n+1)} = \bar{\mathbf{r}}^{(n)} - \Sigma \frac{\bar{\mathbf{r}}^{(n)}}{|\bar{\mathbf{r}}^{(n)}|}\ell_1 {\rm ball}_{\nu_3 {t}_z} (\frac{\bar{\mathbf{r}}^{(n)}}{\Sigma})$
\STATEx \hspace{0.25cm} $\mathbf{b}^{(n+1)} = \bar{\mathbf{b}}^{(n)} - \Sigma \frac{\bar{\mathbf{b}}^{(n)}}{|\bar{\mathbf{b}}^{(n)}|}\ell_1 {\rm ball}_{\nu_4 {t}_B} (\frac{\bar{\mathbf{b}}^{(n)}}{\Sigma})$
\STATEx  \hspace{0.38cm}$\mathbf{t}^{(n+1)} = {\rm neg}({\mathbf{t}^{(n)} + \Sigma \mu \bar{\mathbf{f}}^{(n)}})$
\STATE $\mathbf{f}^{(n+1)} = \mathbf{f}^{(n)}-\mathcal{T}(\mathcal{H}^{\top}\mathbf{w}^{(n+1)}+\nu_1\mathcal{D}_x^{\top}{\mathbf{p}}^{(n+1)} +\nu_2\mathcal{D}_y^{\top}{\mathbf{q}}^{(n+1)}$
\STATEx \hspace{1.3cm} $+ \,\,\nu_3\mathcal{D}_z^{\top}{\mathbf{r}}^{(n+1)}+ \nu_4\mathcal{D}_B^{\top}{\mathbf{b}}^{(n+1)} + \mu \mathbf{t}^{(n+1)})$
 \vspace{0.1cm} 
\STATE $\bar{\mathbf{f}}^{(n+1)} = 2 \mathbf{f}^{(n+1)}-\mathbf{f}^{(n)}$
\STATE $n \leftarrow n+1$
\UNTIL the necessary convergence conditions in Eq. \eqref{eq:convergence_1}are satisfied
\STATE OUTPUT: final SS image $\mathbf{f}^\star=\mathbf{f}^{(n)}$
\end{algorithmic}
\end{algorithm}

In Algorithm \ref{alg:1},  input parameter $\rho\neq 0$ is chosen for empirically optimizing the convergence rate and thus the path of the DTV algorithm leading to the feasible solution set specified by the optimization program in Eq. \eqref{eq:optA}; ${\mathcal{H}}=\mathcal{C}\mathcal{R}\mathcal{D}_B$; $\mathcal{I}$ is an identity matrix of size $I\times I$;  $\mathcal{T}$ and $\Sigma$ are non-diagonal and diagonal matrices, respectively, of size $I\times I$:
\begin{eqnarray}\label{eq:T-sig}
\mathcal{T} &=& \frac{1}{\rho} \left( \frac{\mathcal{I}}{e_L} + \sum_{l=1}^{L-1} \mathbf{u}_l (\frac{1}{e_l} - \frac{1}{e_L})  \mathbf{u}_l^{\top} \right) \quad {\rm and}\quad  \Sigma = \rho \varsigma \mathcal{I},
\end{eqnarray}
$e_l$ and $\mathbf{u}_l$ are the $l$th eigenvalue and eigenvector of matrix $[{\mathcal{H}}^\top {\mathcal{H}}]$, $L$ is the number of eigenvectors  in the expansion and $L=10$ is used in the work; 
$\varsigma={||\mathcal{T} ({\mathcal{K}}^\top {\mathcal{K}})||_2}^{-1}$; $\mathcal{K}^{\top} = ({\mathcal{H}}^{\top}, \nu_1 \mathcal{D}_x^{\top}, \nu_2 \mathcal{D}_y^{\top}, \nu_3 \mathcal{D}_z^{\top}, \nu_4 \mathcal{D}_B^{\top}$, $\mu\mathcal{I}$); 
superscript ``$\top$'' depicts the transpose; $||\cdot||_2$ indicates the largest singular value of a matrix; 
vector $\mathbf{w}^{(n)}$ is of size $J$, whereas vectors $\bar{\mathbf{p}}^{(n)}$, $\bar{\mathbf{q}}^{(n)}$, $\bar{\mathbf{r}}^{(n)}$, $\bar{\mathbf{b}}^{(n)}$, $\mathbf{p}^{(n)}$, $\mathbf{q}^{(n)}$, $\mathbf{r}^{(n)}$, $\mathbf{b}^{(n)}$, and $\mathbf{t}^{(n)}$ are of size $I$; 
operator ${\rm neg}(\cdot)$ enforces the non-positivity constraint; 
operator $\ell_1 {\rm ball}_{\omega}(\cdot)$ projects a vector onto the $\ell_1$-ball of scale $\omega$; $|\bar{\mathbf{q}}|$ depicts a vector of size $I$ with entry $i$ given by $(|\bar{\mathbf{q}}^{(n)}|)_i = |\bar{q}^{(n)}_i|$; and $\bar{q}^{(n)}_i$ indicates the $i$th entry of vector $\bar{\mathbf{q}}^{(n)}$. 
The final reconstruction of SS image $\mathbf{f}^\star$ is obtained when multiple necessary convergence conditions in Eq. \eqref{eq:convergence_1} of the DTV algorithm \cite{zhang2021directional} are reached numerically, as shown in \ref{sec:DTV-convergence-conditions}.

\subsection{Reconstruction evaluation and quantitative metrics}

We conduct visual inspection and quantitative analysis to evaluate the performance of the DTV algorithm in reconstruction of SS images in simulated- and real-data studies.  In visual inspection, we examine the SS-image appearance, spectral profiles, and possible LAR artifacts as compared to their corresponding reference images and spectral profiles, whereas in quantitative analysis, we compute the quantitative metrics described below from SS images reconstructed and compare them with those computed from their reference images. In the simulated-data study, the truth SS image, i.e., the digital phantom, is available and thus is used as the reference image; whereas in the real-data study, we use the SS image obtained from FAR data as the reference image.

The quantitative metrics used in the work include the normalized root-mean-square-error (nRMSE) and Pearson-correlation-coefficient (PCC) \cite{pearson1895notes,Viergever:2003,Bian-PMB:2010,zhang2021directional}, defined as
\begin{eqnarray}
{\rm nRMSE}(\mathbf{f}^{\star}) &=& ||\mathbf{f}^{\star}-\mathbf{f}^{\rm [ref]}||_2/||\mathbf{f}^{\rm [ref]}||_2, \label{eq:imdist}\\
{\rm PCC}(\mathbf{f}^{\star}) & = & \frac{\vert \rm Cov(\mathbf{f}^{\star}, \mathbf{f}^{\rm [ref]})\vert}{\sigma(\mathbf{f}^{\star}) \, \sigma(\mathbf{f}^{\rm [ref]})}, \label{eq:PCC}
\end{eqnarray}
where ${\rm Cov}(\mathbf{f}^{\star}, \mathbf{f}^{\rm [ref]})$ denotes the covariance between final SS image $\mathbf{f}^{\star}$ reconstructed and reference SS image $\mathbf{f}^{\rm [ref]}$; $\sigma^{2}(\mathbf{f})={\rm Cov}(\mathbf{f}, \mathbf{f})$ the variances of $\mathbf{f}$. Metric nRMSE measures the accuracy of a reconstruction relative to the reference, whereas metric PCC yields information about the visual correlation between reconstructed and reference images. We note that ${\rm nRMSE}(\mathbf{f}^{\star}) \rightarrow 0$ and ${\rm PCC}(\mathbf{f}^{\star}) \rightarrow 1$, if $\mathbf{f}^{\star} \rightarrow \mathbf{f}^{\rm [ref]}$. 


Furthermore, spectral profile $V_k(B)$ of contrast material type $k$ at a spatial location in EPRI can be modeled as the convolution of a Gaussian and a Lorentzian, which is  the Voigt function \cite{robinson1999linewidth,mailer2003spectral}, i.e., 
\begin{eqnarray}\label{eq:voigt}
V_k(B) =A_k N_\sigma(B)\circledast L_{\tau_k}(B),
\end{eqnarray}
where $N_\sigma(B)$ denotes a Gaussian with mean $0$ and standard deviation $\sigma$, and 
$L_{\tau_k}(B)=\frac{\tau_k/2}{B^2+(\tau_k/2)^2}$ the Lorentzian of width $\tau_k$, 
where $k=1, 2, ..., K$, indicating type $k$ of contrast material, $K$ is the total number of types of contrast materials within the subject imaged, $\circledast$ denotes convolution, and parameter $A_k$ is determined such that $\int {\rm d}B \, V_k(B)=1$. Lorentzian width $\tau_k$ can be related to oxygen concentration $\eta_k$ within the subject imaged through a linear relationship specified by parameters $l_0$ and $\beta$ as
\begin{eqnarray}\label{eq:line-O2}
\tau_k= l_0 + \beta \times \eta_k.
\end{eqnarray}
We note that parameters $\sigma$, $l_0$, and $\beta$ are independent of oxygen concentration given a contrast material type. In our quantitative evaluation study, we also estimate width $\tau_k$ and oxygen concentration $\eta_k$ from final SS images reconstructed from FAR and LAR data.

\section{Results}\label{sec:results}
While the DTV algorithm in Sec. \ref{sec:DTV_algorithm} can be shown \cite{sidky2012convex,zhang2021directional,chen2021dual} theoretically to solve the convex optimization program in Eq. \eqref{eq:optA}, it remains necessary to verify the correctness of its computer implementation through a numerical study in which measured and model data are consistent, i.e., $\mathbf{g}^{[\mathcal{M}]}=\mathbf{g}$. This can be realized in a simulated-data study in which $\mathbf{g}^{[\mathcal{M}]}$ is  generated by using a digital phantom in Eq. \eqref{eq:dd-model-A}. We first perform a verification study and include the study results in \ref{sec:verification-study} to avoid  distraction from the presentation flow of the main study results of the work below. 

With the DTV algorithm verified, we first carry out a simulated-data study to evaluate the performance of the DTV reconstructions directly from LAR data. Specifically, we use system matrix $\mathcal{H}$ in the DD-data model of Eq. \eqref{eq:dd-model-A} to generate LAR data from the digital phantom and also use $\mathcal{H}$ in Eq. \eqref{eq:optA} for image reconstruction. The purpose of the simulated-data study is to evaluate reconstruction accuracy directly from LAR data relative to the truth SS image for exploring empirically the upper bound of the DTV algorithm's performance in reconstruction of SS images from LAR data. In the real-data study, we apply the DTV algorithm to reconstructing SS images directly from FAR and LAR data collected from a physical phantom, and evaluate the DTV algorithm's performance and robustness with real data containing noise and other physical factors that are inconsistent with (i.e., not included in) the DD-data model in Eq. \eqref{eq:dd-model-A}.

\begin{figure}
\centering
\includegraphics[angle=0,trim=0 0 0 0, clip,origin=c,width=0.8\textwidth]{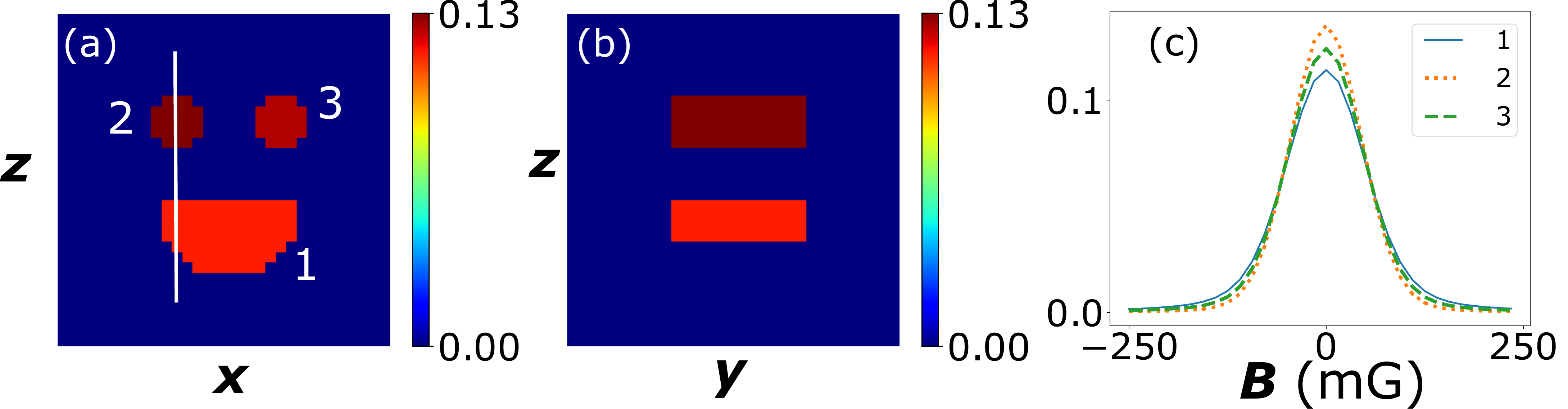}
\caption{The 4D digital phantom (i.e., the truth SS image), which is used in the simulated-data study, within slices specified by $B=0$ G and $y=0$ mm (a) and $B=0$ G and $x=-1.875$ mm (b), respectively, along with the color bars; and (c) the truth spectral profiles at spatial centers of the tubes labeled as $k=1$, 2, and 3 in (a). Display window of [0., 0.13] arbitrary unit (AU) is used in (a) and (b). The image profiles over the white vertical line overlaid in (a) are plotted in Fig. \ref{fig:ss-spatial-lineprofile-simu} below.}
\label{fig:image}
\end{figure}

\subsection{Study materials and conditions}
\subsubsection{Digital and physical phantoms}\label{sec:phantoms} 

In the simulated-data study, we design a digital phantom, i.e., the truth SS image, on a discrete array of $32\times32\times32\times32$ identical voxels each of which is of size $0.31\times 0.31\times 0.31\times0.16$ mm$^3\,$G, covering a 4D support of size $10\!\times \!10\! \times \!10\! \times \!0.5$ mm$^3$G in 4D-SS space $\{x, y, z, B\}$. As shown in Fig. \ref{fig:image}, the digital phantom comprises three ellipsoidal tubes embedded in the 3D-spatial space of zero background. The three tubes contain contrast materials with Voigt functions, specified by Eq. \eqref{eq:voigt}, of different widths $\tau_k$. 
 We select truth $\sigma=44.2$ mG , $l_0=10$ mG, and $\beta=0.5$, as well as truth $\tau_k=$ 10, 22, and 35 mG for tubes $k=$1, 2, and 3. Using truth $\tau_k$s, $l_0$, and $\beta$ in Eq. \eqref{eq:line-O2}, we obtain truth $\eta_k$=0, 24, and 50 torr, respectively, for the three tubes. In Figs. \ref{fig:image}a $\&$ \ref{fig:image}b, we display the 4D phantom (i.e., the truth spectral-spatial image)  within slices specified, respectively, by $B=0$ G \& $y=0$ mm and by $B=0$ G \& $x=-1.875$ mm. Additionally, we plot its truth spectral profile  (i.e., the truth Voigt function) in Fig. \ref{fig:image}c, at the center point within each of the three tubes in Fig. \ref{fig:image}a.

\begin{figure}
\centering
\includegraphics[angle=0,trim=0 0 0 0, clip,origin=c,width=0.85\textwidth]{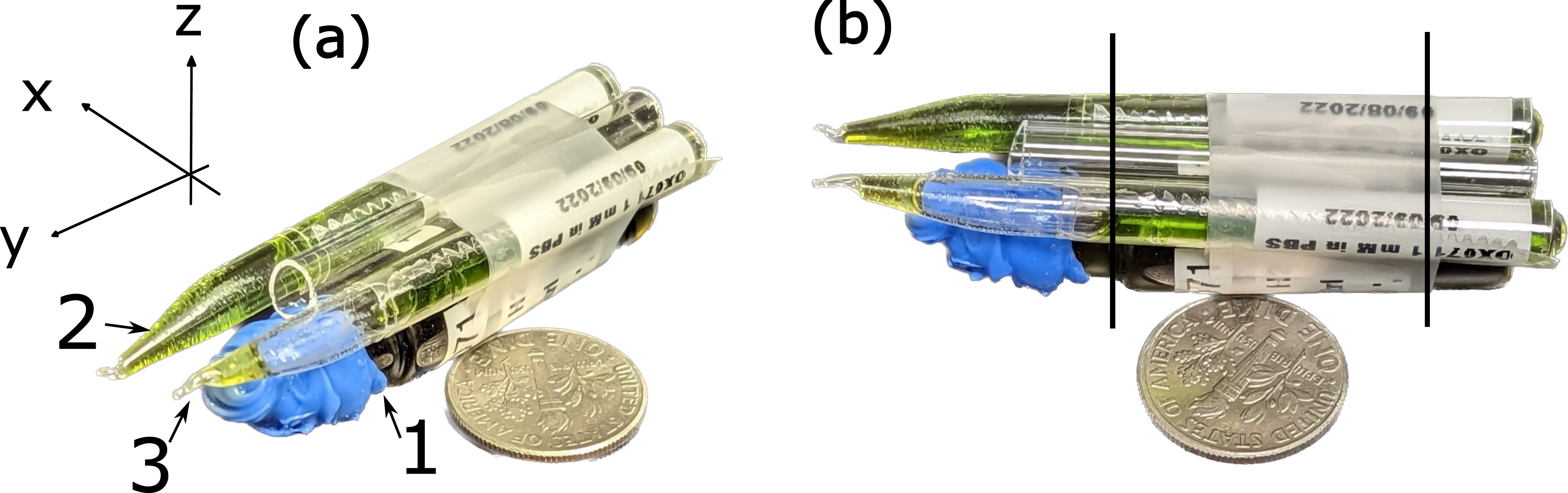}
\caption{The physical phantom  used in the real-data study consists of three borosilicate  glass cylinders with diameters of  9.5 mm, 5 mm, and 5 mm, which are indexed with \#1, \#2, and \#3, respectively, in (a). The physical phantom is placed in a cylindric-shaped resonator of a diameter of 19 mm of our EPRI scanner, and the effective scan region along $y$-axis is indicated between two black lines in (b). A US one-dime coin is placed aside for reference.}
\label{fig:phy-phan}
\end{figure}

In the real-data study, we have fabricated a physical phantom consisting of three borosilicate  glass tubes of cylindrical shape with diameters of  5.0 mm, 5.0 mm, and 9.5 mm, respectively, and identical length of 45 mm, as shown in Fig. \ref{fig:phy-phan}. The bottom glass tube of 9.5 mm diameter is sealed following being filled with 1 mM deoxygenated OX071 trityl solution in PBS, while the top two glass tubes of 5mm diameter are filled with 1 mM OX071 trityl solution in PBS with the addition of different amounts of Cr Maltolate salt to simulate different degrees of oxygen linewidth broadening.
Independent of imaging scans, we have carried out an experiment in which we measured, in addition to $\sigma=31.8$ mG, width $\tau^{(\rm e)}_k$ and oxygen concentration $\eta^{(\rm e)}_k$ within tube $k$, where $k=1, 2, 3$ and superscript ``e'' indicates that they are experimental measurements. Specifically, we obtain the width and oxygen-concentration pairs as ($\tau^{(\rm e)}_1=57.5$ mG, $\eta^{(\rm e)}_1=0$ torr), ($\tau^{(\rm e)}_2=93.5$ mG, $\eta^{(\rm e)}_2=62.24$  torr), and ($\tau^{(\rm e)}_3=135$ mG,  $\eta^{(\rm e)}_3=142.6$ torr), respectively, for tubes $k=$1, 2, and 3. Through fitting the three pairs into Eq. \eqref{eq:line-O2}, we subsequently obtain parameters  $l_0=57.5$ mG and $\beta=0.543$ mG/torr. In our real-data study, we use  $\sigma$, $l_0$, and $\beta$ determined experimentally in our estimation of $\tau_k$ and $\eta_k$ from 4D images reconstructed from FAR or LAR data, as presented below.

\subsubsection{FAR and LAR scans}\label{sec:data-generation}

\begin{table}
\caption{FAR and LAR scans considered in the simulated-data study.}
\centering
\begin{tabular}{|c|c|c|c|}\hline
\diagbox[linewidth=1.5pt,width=20mm,height=10mm]{$\gamma_\tau$}{($\theta_\tau$, $\phi_\tau$)} & (85.5$^\circ$, 90$^\circ$) & (85.5$^\circ$, 45$^\circ$) & (67.5$^\circ$, 45$^\circ$) \\ \hline
85.5$^\circ$ & FAR   & LAR1   & LAR2  \\ \hline
67.5$^\circ$ & LAR3  & LAR4   & LAR5  \\ \hline
58.5$^\circ$ & LAR6  & LAR7   & LAR8  \\ \hline
\end{tabular}
\label{table:LAR-views-simu}
\end{table}

In the simulated-data study, for the FAR scan specified by $\xi_{\tau} = 250$ mG, $\gamma_{\tau}=90^\circ$, $\theta_{\tau}=90^\circ$, and $\phi_{\tau}=90^\circ$, its ESA-sampling grids are determined by use of intervals $\Delta_\xi=7.8$ mG,  $\Delta_\gamma=9^\circ$, $\Delta_\theta=9^\circ$, and $\Delta_\phi(\theta_n)={\rm min}(|\frac{\Delta_\gamma}{{\rm sin} (\theta_n)}|,\pi)$, where $\theta_n$ denotes the $n$th sample on $\theta$, as discussed in Sec. \ref{sec:data-acquisition}. 
As discussed above, it is physically impossible to acquire data near the vicinity of $\gamma_\tau= 90^\circ$. Therefore, an effective FAR scan is specified in Table \ref{table:LAR-views-simu} that has $\gamma_\tau= 85.5^\circ$.  Using the digital phantom of Fig. \ref{fig:image} in Eq. \eqref{eq:dd-model-A}, we generate FAR data on the ESA-sampling grids. For a LAR scan specified in Table \ref{table:LAR-views-simu}, we obtain the LAR data by choosing a subset of the FAR data in the portion of the FAR scan corresponding to the angular ranges of the LAR scan.

\begin{table}
\caption{FAR and LAR scans considered in the real-data study.}
\centering
\begin{tabular}{|c|c|c|c|}\hline
\diagbox[linewidth=1.5pt,width=20mm,height=10mm]{$\gamma_\tau$}{($\theta_\tau$, $\phi_\tau$)} & (84.4$^\circ$, 90$^\circ$) & (84.4$^\circ$, 45$^\circ$) & (73.1$^\circ$, 45$^\circ$) \\ \hline
84.4$^\circ$ & FAR & LAR1  & LAR2   \\ \hline
73.1$^\circ$ & LAR3   & LAR4   & LAR5     \\ \hline
61.9$^\circ$ & LAR6   & LAR7   & LAR8    \\ \hline
\end{tabular}
\label{table:LAR-views-real}
\end{table}


In the real-data study,  we utilized a spectroscopic imager \cite{halpern1989imaging,ahn2007spatially} to scan the phantom operating at radio frequency (RF) of 250 MHz with a non-saturated level of RF power of 40 $\mu$W. During the scan, we applied a 5 kHz Zeeman modulation with an amplitude of 60 mG. The strength of the magnetic field gradient ranges from -350 mG/mm to 350 mG/mm.
A total of 2624 sweeps were obtained for the FAR scan, with each sweep acquired at 256 field points along the $\xi$-axis and a scan time of 3 ms per point.

For the FAR scan specified by $\xi_{\tau} = 724$ mG, $\gamma_{\tau}=90^\circ$, $\theta_{\tau}=90^\circ$, and $\phi_{\tau}=90^\circ$, its ESA-sampling grids are determined by use of $\Delta_\xi=5.66$ mG,  $\Delta_\gamma=11.25^\circ$, $\Delta_\theta=11.25^\circ$, and $\Delta_\phi(\theta_n)={\rm min}(|\frac{\Delta_\gamma}{{\rm sin} (\theta_n)}|,\pi)$, where $\theta_n$ denotes the $n$th sample on $\theta$, as discussed in Sec. \ref{sec:data-acquisition}. For the same reason discussed in the simulated-data study, an effective FAR scan is specified in Table \ref{table:LAR-views-real} that has $\gamma_\tau = 84.4^\circ$.  
Using our EPRI system, we collect FAR data from the physical phantom in  Fig. \ref{fig:phy-phan}. For a LAR scan specified in Table \ref{table:LAR-views-real}, we obtain the LAR data by choosing a subset of the FAR data in the portion of the FAR corresponding to the angular ranges of the LAR scan.

\subsubsection{Reconstruction parameters}
In the simulated-data study, SS images are reconstructed on image arrays identical to that of the truth SS image. Values of constraint parameters ($t_x$, $t_y$, $t_z$, $t_B$) are chosen to be the DTVs computed, respectively, along $x$, $y$, $z$, and $B$ dimensions of the truth SS image, i.e., the known digital phantom in the simulated-data study. In an attempt to streamline the presentation flow, we opt not to include results of a noisy simulated-data study, because we also perform in the work a real-data study in which data measured contain not only noise but also other physical factors inconsistent with the DD-data model Eq. \eqref{eq:dd-model-A} upon which the DTV algorithm is developed, and all of these physical factors may impact SS images reconstructed. 

In the real-data study,  from each set of data collected in FAR and LAR scans specified in Table \ref{table:LAR-views-real}, we perform reconstructions of SS images by using the DTV algorithm for multiple sets of values of constraint parameters ($t_x$, $t_y$, $t_z$, $t_B$) and then  select the set that yields an SS image with visually minimal artifacts. The SS images are reconstructed on image arrays of $64\times64\times64\times64$ identical voxels each of which is of size {\color{black} $0.66\times 0.66\times 0.66\times22.6$} mm$^3\,$mG, covering a 4D support of size {\color{black} $42\!\times \!42\! \times \!42\! \times \!1448$} mm$^3$mG in 4D-SS space $\{x, y, z, B\}$.

\subsection{SS images reconstructed in the simulated-data study}\label{sec:recon-SS-images}

The DTV algorithm is applied to reconstructing SS images from data of the digital phantom for the FAR and LAR scans summarized in Table \ref{table:LAR-views-simu} above.

\subsubsection{Reconstruction results}

\begin{figure}
\centering
\includegraphics[angle=0,trim=0 0 0 0, clip,origin=c,width=0.95\textwidth]{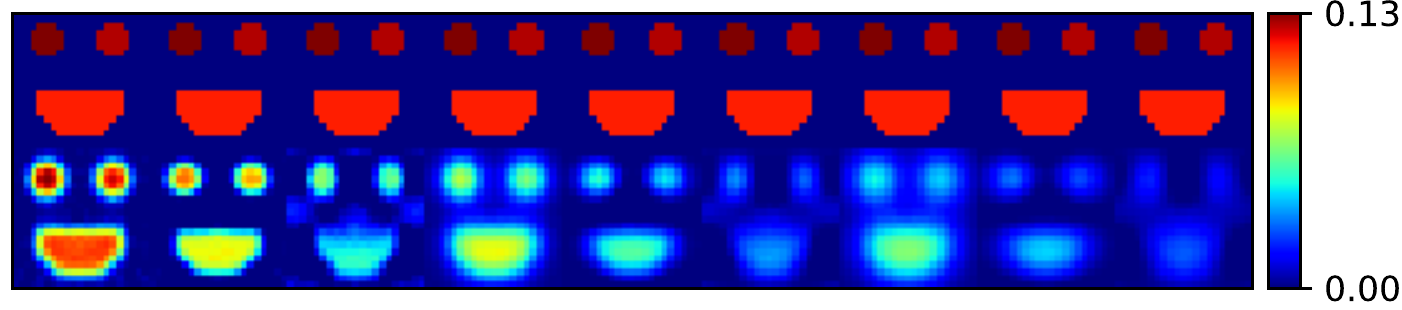}
\caption{SS images within slice $B=0$ G and $y = 0$ mm reconstructed by use of the DTV algorithm (row 1) and FBP algorithm (row 2) from simulated data of the FAR scan (column 1) and LAR scans LAR1-LAR8 (columns 2-9) summarized in Table \ref{table:LAR-views-simu}, respectively. Display window: [0., 0.13] AU.}
\label{fig:ss-trans32-simu}
\end{figure}

\begin{figure}
\centering
\includegraphics[angle=0,trim=0 0 0 0, clip,origin=c,width=0.95\textwidth]{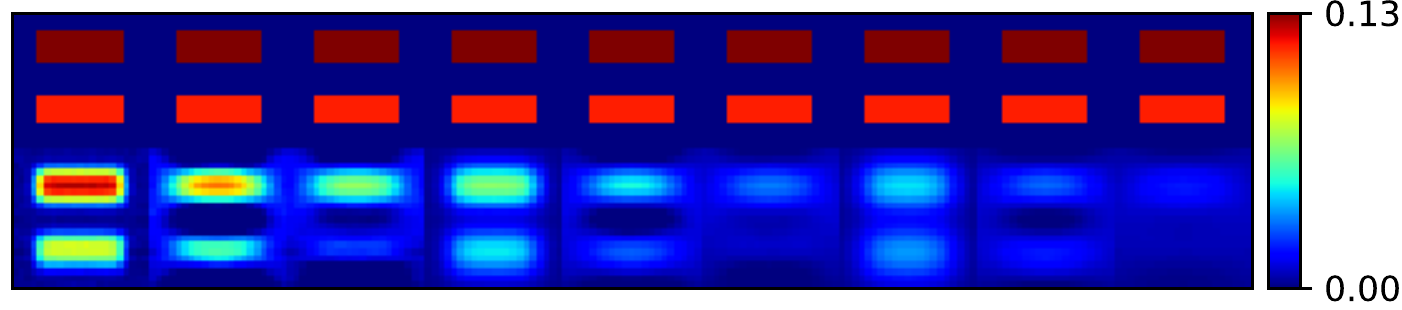}
\caption{SS images within slice $B=0$ G and $x = -1.875$ mm reconstructed by use of the DTV algorithm (row 1) and FBP algorithm (row 2) from simulated  data of the FAR scan (column 1) and LAR scans LAR1-LAR8 (columns 2-9) summarized in Table \ref{table:LAR-views-simu}, respectively. Display window: [0., 0.13] AU.}
\label{fig:ss-sag32-simu}
\end{figure}

We first perform reconstruction from FAR data and display the SS images reconstructed within the $x$-$y$ plane specified by $B=0$ G \& $y=0$ mm and the $y$-$z$ plane specified by $B=0$ G \& $x=-1.875$ mm, respectively, in Figs. \ref{fig:ss-trans32-simu} and \ref{fig:ss-sag32-simu}. 
The result verifies that the DTV algorithm and its computer implementation can solve the optimization program in Eq. \eqref{eq:optA} and accurately reconstruct SS images from FAR data. 

We subsequently apply the DTV algorithm verified to reconstructing SS images from data of LAR scans specified in Table \ref{table:LAR-views-simu}. The reconstructed SS images within the $x$-$y$ plane specified by $B=0$ G \& $y=0$ mm and the $y$-$z$ plane specified by $B=0$ G \& $x=-1.875$ mm are shown, respectively, in Figs. \ref{fig:ss-trans32-simu} and \ref{fig:ss-sag32-simu}. For reference, we  also perform FBP reconstructions from FAR and LAR data and display them in the bottom rows of Figs. \ref{fig:ss-trans32-simu} and \ref{fig:ss-sag32-simu}.

It can be observed that the DTV algorithm reconstructs accurately SS images from data of LAR scans depicted in Table \ref{table:LAR-views-simu}, as the SS images appear to be largely free of LAR artifacts and visually resemble the truth SS image. 
On the other hand, the FBP images reconstructed from LAR data appear to suffer considerably from artifacts that distort the anatomic structures in, and introduce biases to, the SS images. Also, while the SS images reconstructed from FAR data by use of the DTV and FBP algorithms appear somewhat comparable, the former is visually free of minor artifacts observable in the latter.

\begin{figure}
\centering
\includegraphics[angle=0,trim=0 0 0 0, clip,origin=c,width=0.95\textwidth]{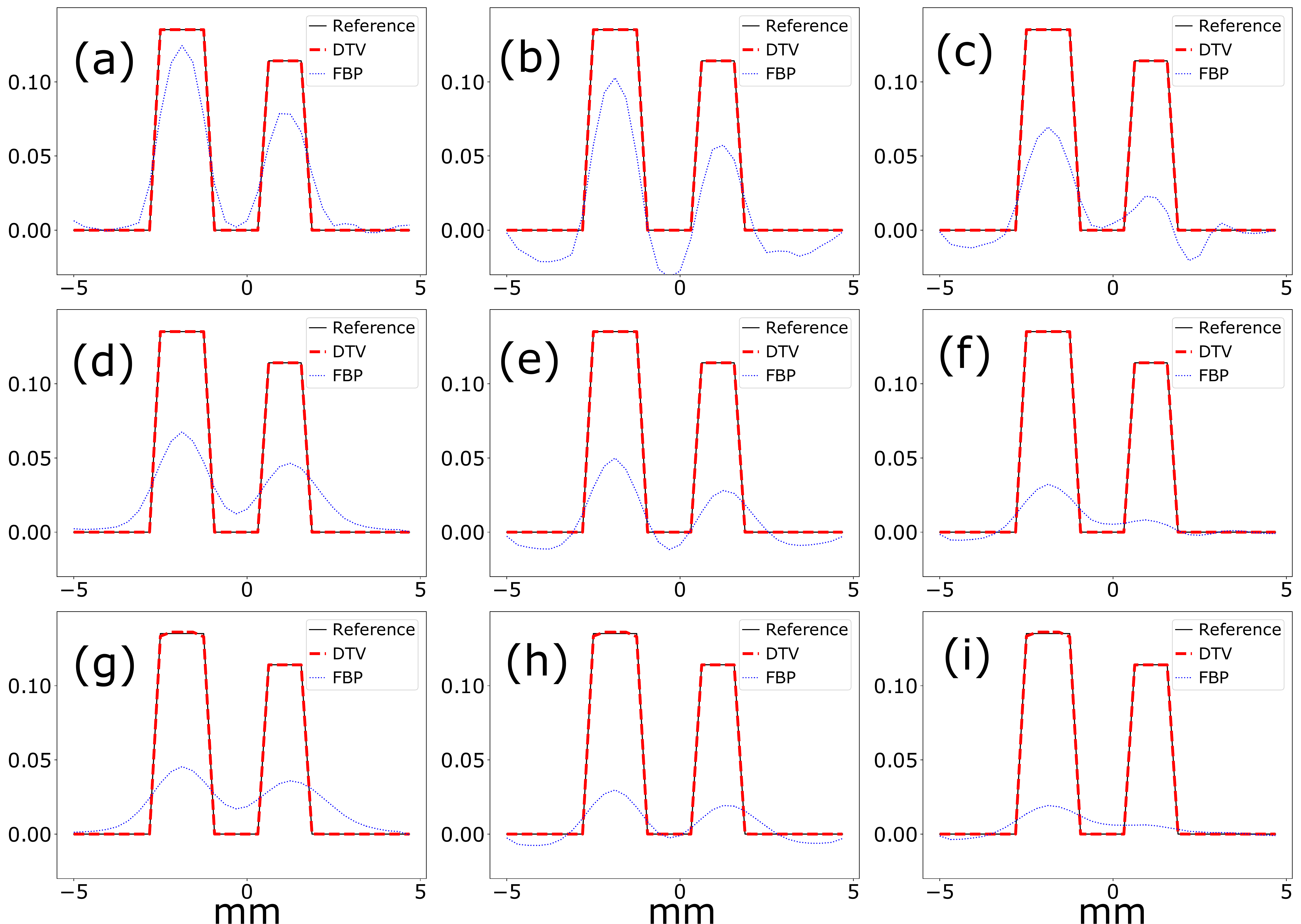}
\caption{Spatial profiles (over the white line overlaid in the SS image in Fig. \ref{fig:image}a) of the reference (i.e., the truth) SS image (solid), and of the SS images reconstructed by use of the DTV algorithm (dashed) and the FBP algorithm (dotted) from simulated data of FAR scan (a) and LAR scans LAR1-LAR8 (b-i) depicted in Table \ref{table:LAR-views-simu}.
Similar results of spatial profiles are obtained also over other lines in SS images reconstructed from data of LAR scans depicted in  Table \ref{table:LAR-views-simu}, and thus are not shown.}
\label{fig:ss-spatial-lineprofile-simu}
\end{figure}

\begin{figure}
\centering
\includegraphics[angle=0,trim=0 0 0 0, clip,origin=c,width=0.95\textwidth]{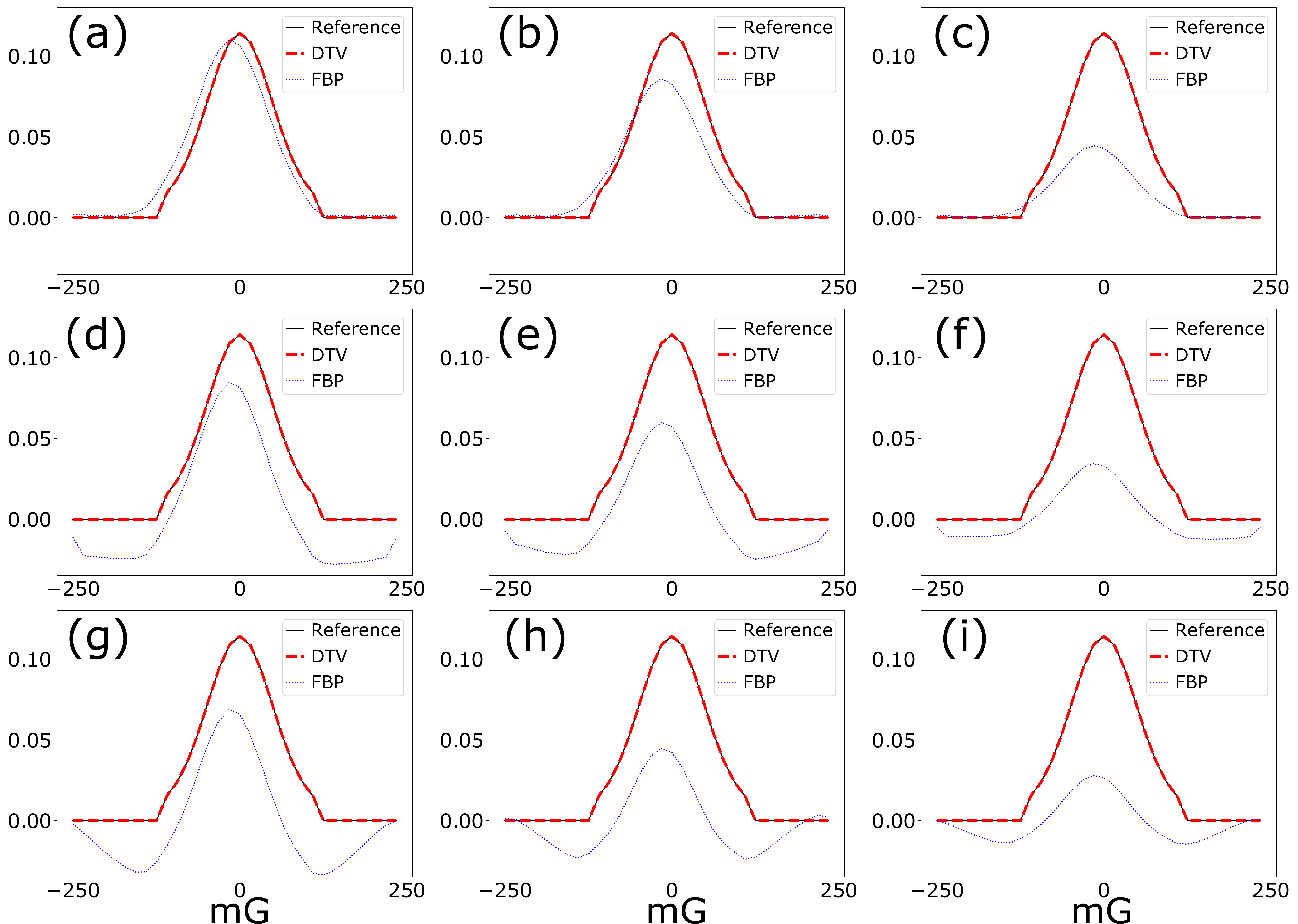}
\caption{Spectral profiles (at a spatial point in tube 1 specified by $x=0$ mm, $y=0$ mm, and $z=-1.24$ mm) of the reference (i.e., the truth) SS image (solid), and of the SS images reconstructed by use of the DTV algorithm (dashed) and the FBP algorithm (dotted) from simulated data of FAR scan (a) and LAR scans LAR1-LAR8 (b-i) depicted in  Table \ref{table:LAR-views-simu}.
Similar results are obtained for spectra at other spatial locations in SS images reconstructed from data of LAR scans depicted in Table \ref{table:LAR-views-simu} and thus are not shown.}
\label{fig:ss-spectral-lineprofile-simu}
\end{figure}

For revealing reconstruction accuracy, we also plot in Figs. \ref{fig:ss-spatial-lineprofile-simu} and \ref{fig:ss-spectral-lineprofile-simu} the spatial profiles (over the white line in Fig. \ref{fig:image}a) and the spectral profile at a spatial point of $x=0$ mm, $y=0$ mm, and $z=1.24$ mm (in tube 1) in SS images reconstructed from FAR and LAR data. The spatial and spectral profiles in the SS images reconstructed by use of the DTV algorithm  appear to be in good agreement with their counterparts in the truth SS image, whereas those in the FBP reconstructions differ significantly from that in the truth SS image.

\subsubsection{Quantitative evaluation}\label{sec:evaluation-simu}

\begin{figure}
\centering
\includegraphics[angle=0,trim=0 0 0 0, clip,origin=c,width=0.95\textwidth]{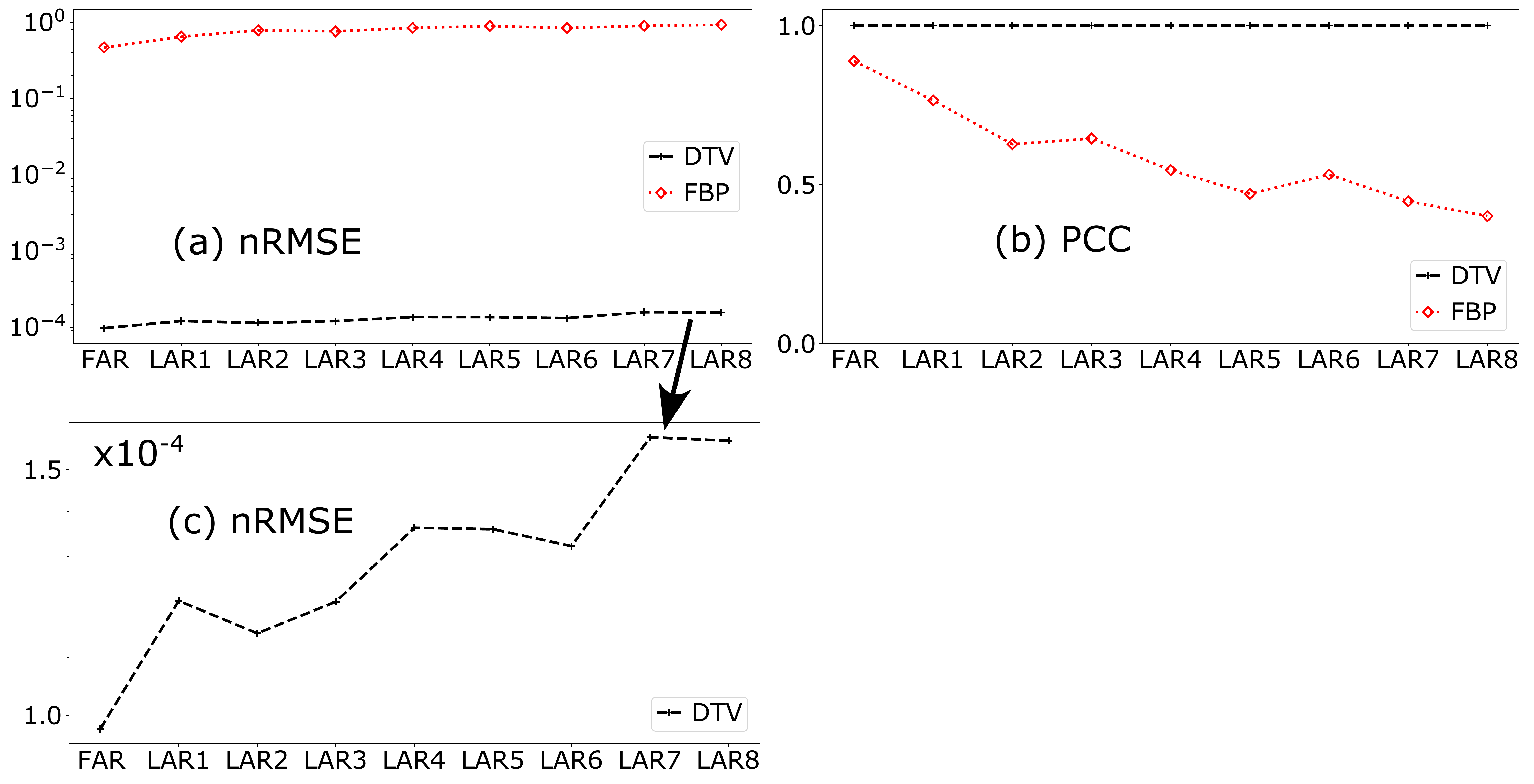}
\caption{Quantitative metrics nRMSE (a) and PCC (b) of the SS images reconstructed by use of the DTV (black, dashed) and FBP (red, dotted) algorithms from simulated data of FAR and LAR scans specified in Table. \ref{table:LAR-views-simu}. A zoomed-in view of nRMSEs of DTV reconstructions are shown in (c).}
\label{fig:ss-metrics-simu}
\end{figure}

Using metrics nRMSE and PCC in Eqs. \eqref{eq:imdist} and \eqref{eq:PCC}, we quantitatively evaluate reconstruction accuracy of SS images relative to the reference image, i.e., the truth SS image. In Fig. \ref{fig:ss-metrics-simu}, we show the nRMSEs and PCCs computed from the SS images reconstructed by using the DTV algorithm directly from data of the FAR and LAR scans specified in Table \ref{table:LAR-views-simu}. It can be observed that, as the angular ranges of LAR scans increase, the nRMSE decreases while the PCC remains largely 1, corroborating the observation above that the SS images reconstructed appear visually comparable to the truth SS image. We also calculate metrics nRMSE and PCC from the FBP reconstructions, which are shown in Fig. \ref{fig:ss-metrics-simu}. We observe that the nRMSEs of the FBP reconstructions are significantly higher than those of the DTV reconstructions, while the PCCs of the FBP reconstructions are considerably lower than 1.

\begin{figure}
\centering
\includegraphics[angle=0,trim=0 0 0 0, clip,origin=c,width=0.95\textwidth]{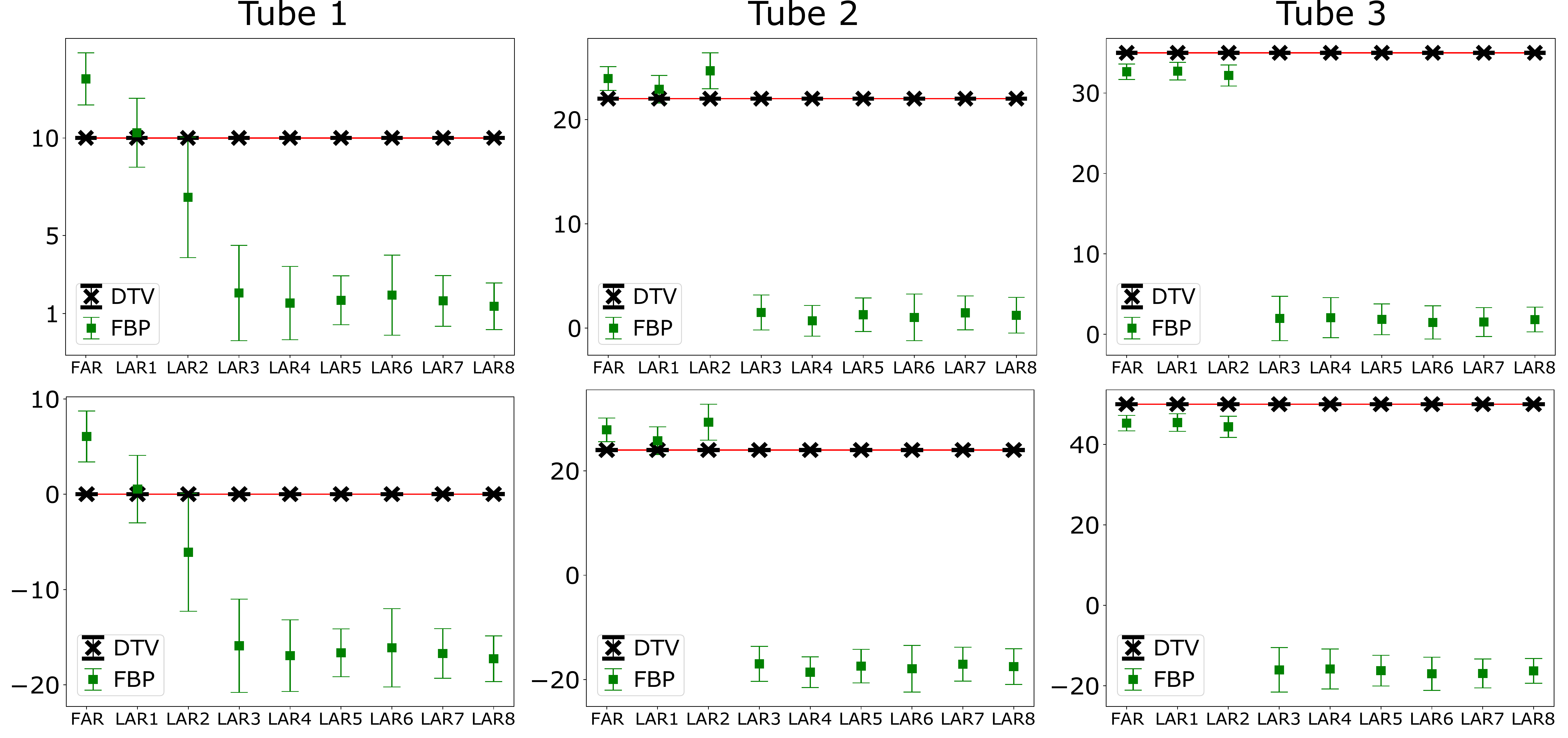}
\caption{Mean widths $\bar{\tau}_k$ (row 1) and mean oxygen-concentrations $\bar{\eta}_k$ (row 2), along with their error bars (i.e., $\sigma_k^{\tau}$s and  $\sigma_k^{\eta}$s), where $k=1, 2,$ and 3 (columns 1-3) within the 3 tubes of the numerical phantom in Fig. \ref{fig:image} estimated, respectively, from the DTV (black, `$\boldsymbol{\times}$') and the FBP (green, `{\color{green}$\blacksquare$}') reconstructions from simulated data of FAR and LAR scans specified in Table. \ref{table:LAR-views-simu}. Solid red line indicates the corresponding reference (i.e., truth) value.}
\label{fig:ss-parameters-simu}
\end{figure}

We also fit the Voigt function with the spectral profile at a selected 3D-spatial voxel within an SS image reconstructed by using the DTV or FBP algorithm to estimate \cite{robinson1999linewidth,mailer2003spectral} the Lorentzian line width $\tau_k$, which is subsequently used to compute the oxygen concentration in the solution by use of Eq. \eqref{eq:line-O2}. Specifically, we select three 3D-spatial regions of interest (ROIs), and each contains 90 3D-voxels, respectively, at the centers of tubes 1, 2, and 3. 
For each ROI, we first estimate $\tau_k$ at each of its voxels by fitting the spectral profile to a Voigt curve
 and then calculate spatial mean $\bar{\tau}_k$ and (spatial) standard deviation $\sigma_k^{\tau}$ by averaging $\tau_k$s estimated over the voxels within the ROI, which are shown in row 1 of Fig. \ref{fig:ss-parameters-simu}.  Using $\bar{\tau}_k$ in Eq. \eqref{eq:line-O2}, along with truth $l_0$ and $\beta$, we calculate spatial mean $\bar{\eta}_k$. Similarly, (spatial) standard deviation $\sigma_k^{\eta}$ can readily be computed by using $\sigma_k^{\tau}$ in Eq. \eqref{eq:line-O2}. It can be observed that $\bar{\tau}_k$ and $\bar{\eta}_k$ estimated  from reconstructions from LAR data by use of  the DTV algorithm agree well with the corresponding truth values. On the other hand,  $\bar{\tau}_k$ and $\bar{\eta}_k$ estimated 
  from the FBP reconstructions appear to deviate considerably from the truth values.

\subsection{SS images reconstructed in the real-data study}\label{sec:results-real-data}

\subsubsection{Reconstruction results}
\begin{figure}
\centering
\includegraphics[angle=0,trim=0 0 0 0, clip,origin=c,width=1\textwidth]{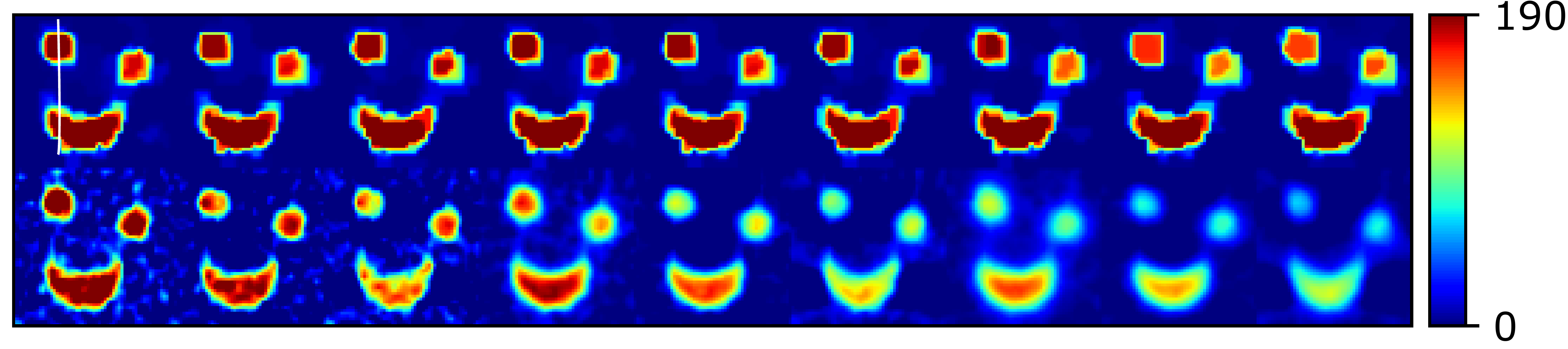}
\caption{SS images within slice $B=0$ G and {\color{black} $y = -3.27$} mm reconstructed by use of the DTV algorithm (row 1) and FBP algorithm (row 2) from real data of the FAR scan (column 1) and LAR scans LAR1-LAR8 (columns 2-9) summarized in Table \ref{table:LAR-views-real}, respectively. Display window: [0, 190] AU. The image profiles over the white vertical line overlaid on the image (row 1 and column 1) are plotted in Fig. \ref{fig:ss-spatial-lineprofile-real} below.}
\label{fig:ss-trans32-real}
\end{figure}

\begin{figure}
\centering
\includegraphics[angle=0,trim=0 0 0 0, clip,origin=c,width=1\textwidth]{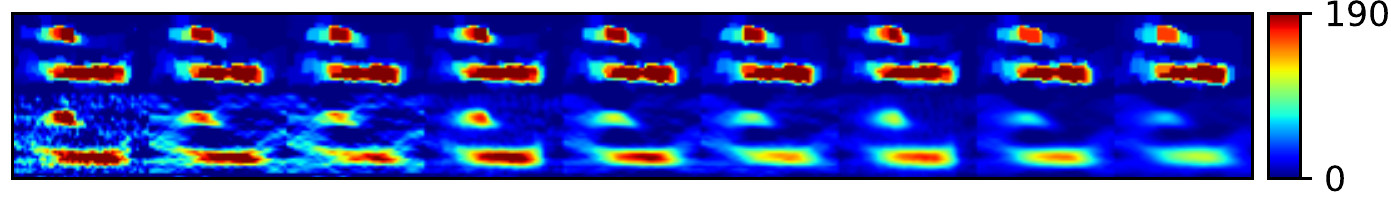}
\caption{SS images within slice $B=0$ G and {\color{black} $x= -3.93$} mm reconstructed by use of the DTV algorithm (row 1) and FBP algorithm (row 2) from real data of the FAR scan (column 1) and LAR scans LAR1-LAR8 (columns 2-9) summarized in Table \ref{table:LAR-views-real}, respectively. Display window: [0, 190] AU.}
\label{fig:ss-sag32-real}
\end{figure}

Using our EPRI system \cite{epel2011comparison}, we acquire data from the physical phantom, shown in Fig. \ref{fig:phy-phan}, for the FAR and LAR scans summarized in Table \ref{table:LAR-views-real} above.
We then apply the DTV algorithm to reconstructing SS images from data acquired. For each of the FAR and LAR scans in Table \ref{table:LAR-views-real}, we first carry out reconstructions  with multiple sets of constraint-parameter values chosen, and subsequently select the optimal set that yields the SS image with visually minimal artifacts.  In Figs. \ref{fig:ss-trans32-real} and \ref{fig:ss-sag32-real}, we display the SS images reconstructed within the $x$-$z$ plane specified by $B=0$ G \& {\color{black} $y = -3.27$} mm and the $x$-$y$ plane specified by $B=0$ G \& {\color{black} $x= -3.93$} mm, respectively, each of which is obtained when its corresponding optimal set of constraint-parameter values selected is achieved numerically \cite{zhang2021directional}.

It can be observed that the DTV algorithm reconstructs SS images directly from LAR data with minimal LAR artifacts and visually resemble the reference SS image reconstructed from FAR data. Again, as the angular ranges of LAR scans further decrease, the visual quality of DTV reconstructions of the SS image appears to be reduced slightly especially for LAR scans with smallest angular ranges in Table \ref{table:LAR-views-real}. However, the FBP images reconstructed from LAR data appear to be with considerable artifacts that distort the anatomic structures in, and introduce biases to, the SS images. Also, the SS image reconstructed from FAR data by use of the DTV algorithm appears visually free of much of the artifacts observable in the FBP image.

\begin{figure}
\centering
\includegraphics[angle=0,trim=0 0 0 0, clip,origin=c,width=0.95\textwidth]{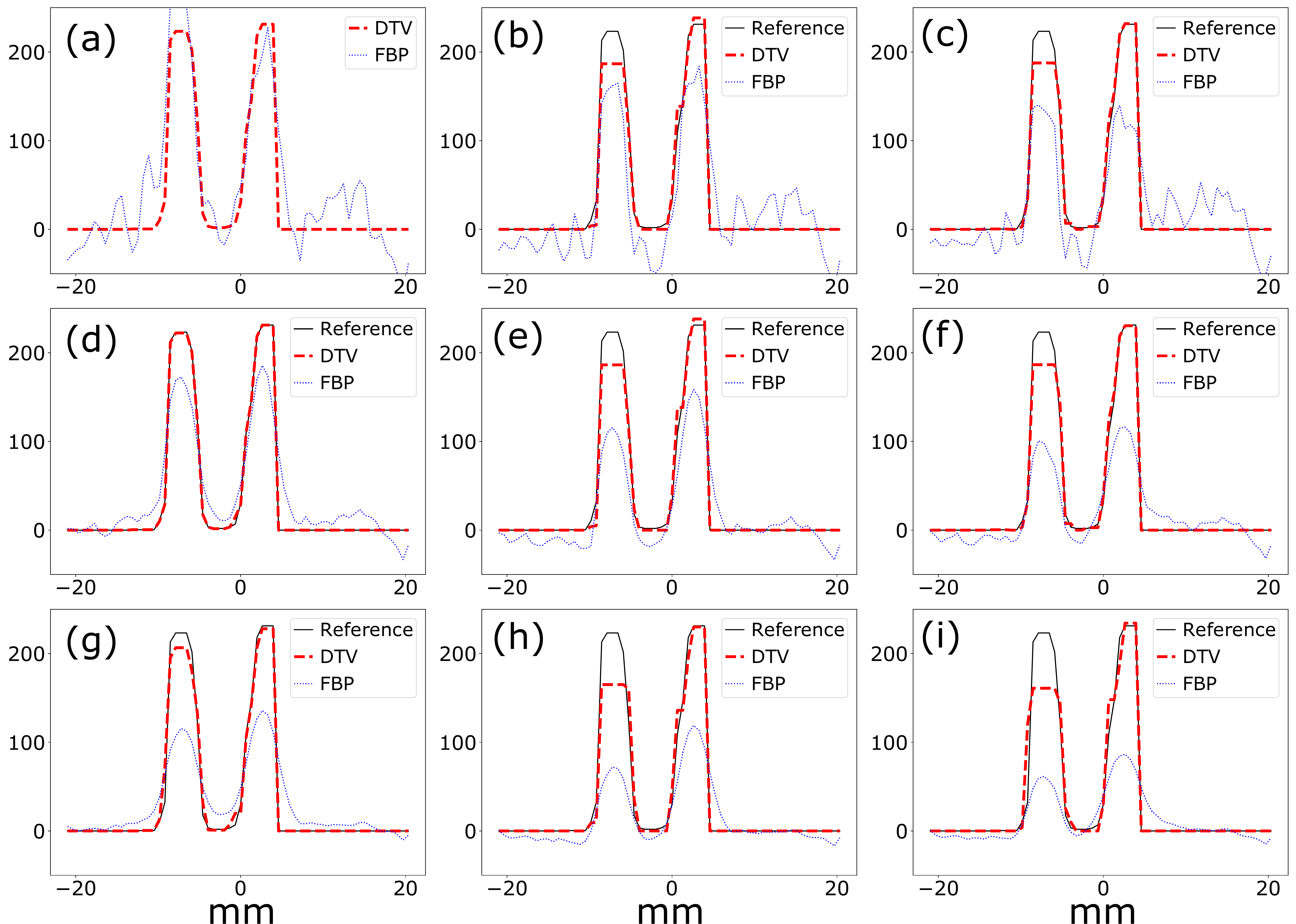}
\caption{Spatial profiles (over the white line overlaid in the SS image in the top left panel of Fig. \ref{fig:ss-trans32-real}) of the reference SS image (solid) reconstructed from the FAR data by use of the DTV algorithm, and of the SS images reconstructed by use of the DTV algorithm (dashed) and the FBP algorithm (dotted) from real data of FAR scan (a) and LAR scans LAR1-LAR8 (b-i) depicted in Table \ref{table:LAR-views-real}.
Similar results are obtained for spatial profiles over other lines in SS images reconstructed from data of the FAR and LAR scans depicted in Table \ref{table:LAR-views-real} and thus are not shown.}
\label{fig:ss-spatial-lineprofile-real}
\end{figure}

\begin{figure}
\centering
\includegraphics[angle=0,trim=0 0 0 0, clip,origin=c,width=0.95\textwidth]{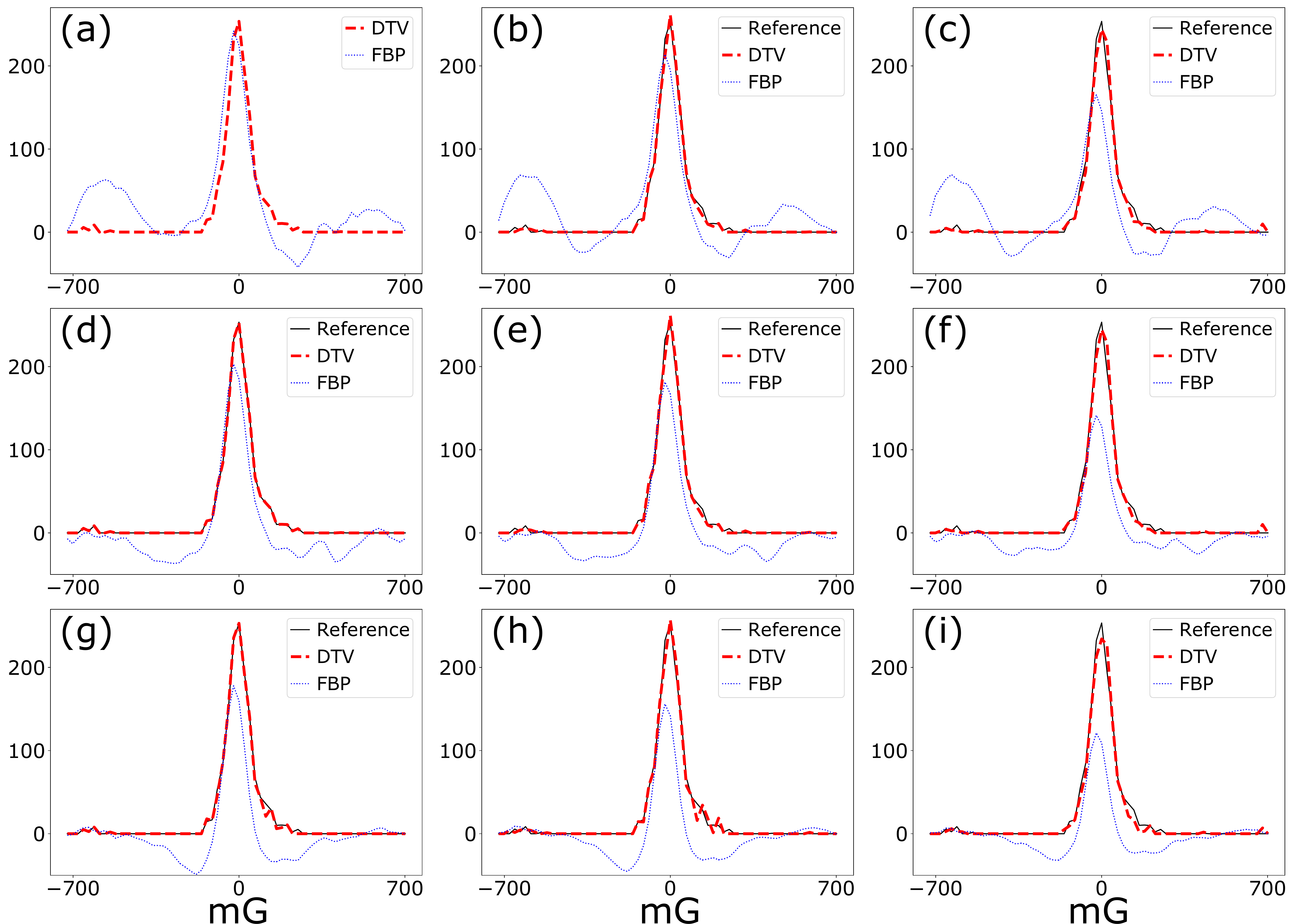}
\caption{Spectral profiles (at a spatial point in glass tube 1 specified by {\color{black} $x=0.66$ mm, $y=-3.27$ mm, and $z=-3.27$} mm) of the reference SS image (solid) reconstructed from the FAR data by use of the DTV algorithm, and of the SS images reconstructed by use of the DTV algorithm (dashed) and the FBP algorithm (dotted) from real data of FAR scan (a) and LAR scans LAR1-LAR8 (b-i) depicted in  Table \ref{table:LAR-views-real}.
Similar results are obtained for spectra at other spatial locations in SS images reconstructed from data of LAR scans depicted in Table \ref{table:LAR-views-real} and thus are not shown.}
\label{fig:ss-spectral-lineprofile-real}
\end{figure}

In addition, we plot in Figs. \ref{fig:ss-spatial-lineprofile-real} and \ref{fig:ss-spectral-lineprofile-real}  the spatial profiles over the white line in Fig. \ref{fig:ss-trans32-real}a and spectral profile for a spatial point specified by {\color{black} $x=0.66$ mm, $y=-3.27$  mm, and $z=-3.27$} mm in SS images reconstructed from FAR and LAR data. The spatial and spectral profiles in DTV reconstructions appear to be in agreement with their counterparts in the reference SS image reconstructed from FAR data by use of the DTV algorithm, whereas those in the FBP reconstructions differ significantly from that in the reference SS image.

\subsubsection{Quantitative Evaluation}\label{sec:evaluation-real}

\begin{figure}
\centering
\includegraphics[angle=0,trim=0 0 0 0, clip,origin=c,width=0.95\textwidth]{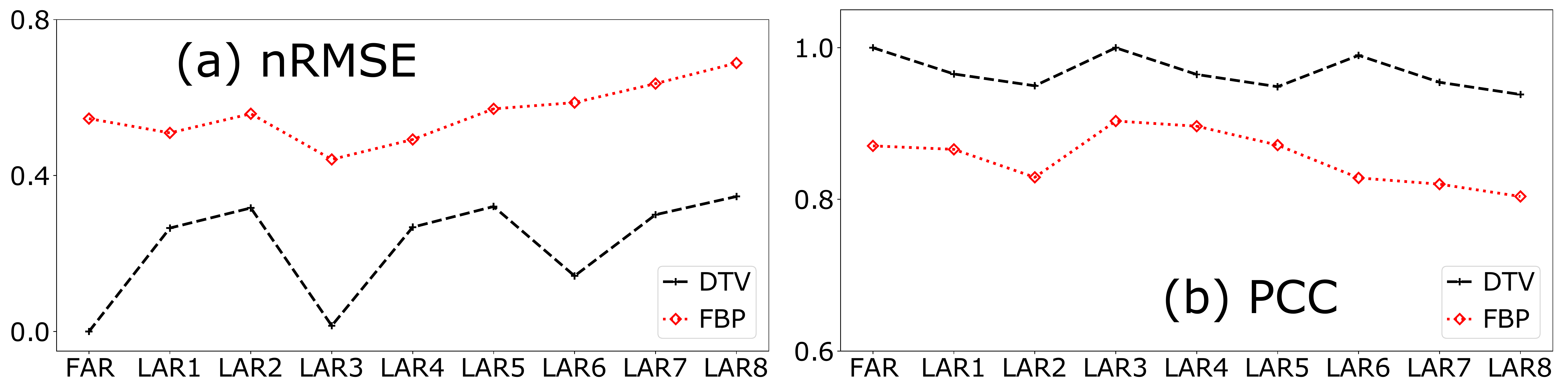}
\caption{Quantitative metrics nRMSE (a) and PCC (b) of the SS images reconstructed by use of the DTV (black, dashed) and FBP (red, dotted) algorithms from real data of FAR and LAR scans specified in Table. \ref{table:LAR-views-real}.}
\label{fig:ss-metrics-real}
\end{figure}

Using metrics nRMSE and PCC in Eqs. \eqref{eq:imdist} and \eqref{eq:PCC}, we also perform a quantitative evaluation of reconstruction accuracy relative to the reference image, i.e., the SS image reconstructed from FAR data by use of the DTV algorithm. In Fig. \ref{fig:ss-metrics-real}, we show nRMSEs and PCCs computed from the SS images reconstructed by using the DTV algorithm directly from data of the LAR scans specified in Table \ref{table:LAR-views-real}. It can be observed that, as the angular ranges of a LAR scan increase, the nRMSE decreases while PCC remains largely 1, corroborating the observation above that the SS images reconstructed appear visually comparable to the reference SS image. We also show in Fig. \ref{fig:ss-metrics-real} nRMSE and PCC calculated from the FBP reconstructions, and it can be observed that the nRMSE and PCC are considerably higher and lower, respectively, than  those of the DTV reconstructions. 

In the real-data study, we have $\sigma=31.8$ mG, $l_0=57.5$ mG, and $\beta=0.543$ mG/torr, as discussed in Sec. \ref{sec:phantoms} above.
For a 4D spectral-spatial image reconstructed from FAR or LAR data by using the DTV or FBP algorithm, we choose three 3D-spatial ROIs containing 72, 48, and 48 3D-voxels, respectively, in glass tubes $k=1$, 2, and 3. For each ROI, we first estimate $\tau_k$ at each of its voxels by fitting the spectral profile to the Voigt function with $\sigma=31.8$ mG; and we then compute  spatial mean $\bar{\tau}_k$, and corresponding (spatial) standard deviation $\sigma_k^{\tau}$, from $\tau_k$s  estimated within the ROI.  In particular, for $\bar{\tau}_k$ of images reconstructed from FAR data by using each of the DTV and FBP algorithms, we obtain a calibration coefficient by matching $\bar{\tau}_k$ with the corresponding $\tau^{(\rm e)}_k$ determined in Sec. \ref{sec:phantoms} above. Subsequently, applying the calibration coefficient determined for the DTV or FBP algorithm to its corresponding reconstructions from LAR data, we obtain $\bar{\tau}_k$s  calibrated for the LAR scans. Using each of these  $\bar{\tau}_k$s in Eq. \eqref{eq:line-O2}, along with $l_0=57.5$ mG and $\beta=0.543$ mG/torr, we then obtain $\bar{\eta}_k$ for FAR and LAR scans. Similarly, we can readily obtain $\sigma_k^{\eta}$ by using its corresponding $\sigma_k^{\tau}$ in Eq. \eqref{eq:line-O2}.

\begin{figure}
\centering
\includegraphics[angle=0,trim=0 0 0 0, clip,origin=c,width=0.95\textwidth]{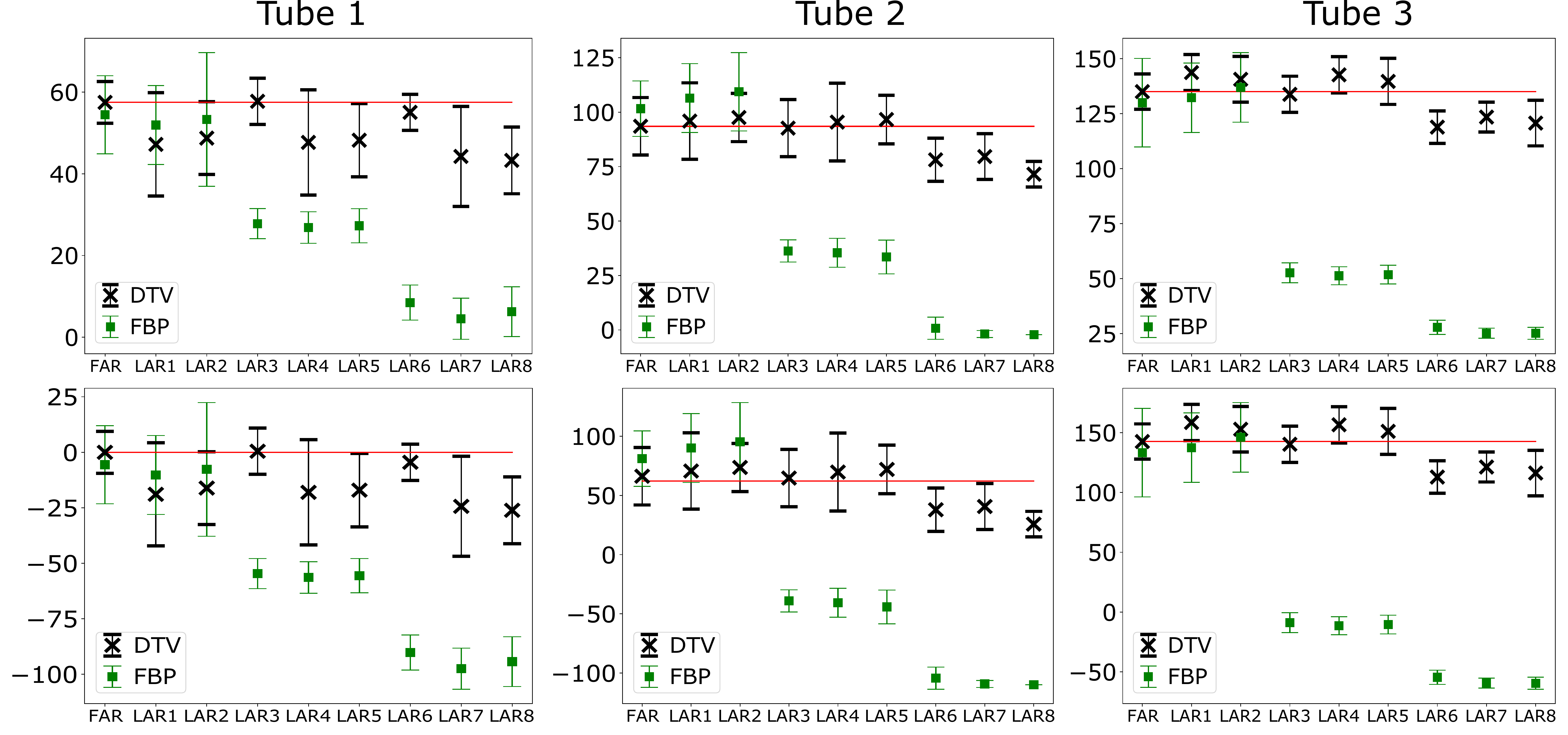}
\caption{Mean widths $\bar{\tau}_k$ (row 1) and oxygen concentrations $\bar{\eta}_k$ (row 2), along with their error bars (i.e., $\sigma_k^{\tau}$ and  $\sigma_k^{\eta}$), where $k=1, 2,$ and 3 (columns 1-3) within the 3 glass tubes of the physical phantom in Fig. \ref{fig:phy-phan} estimated, respectively, from the DTV (black, `$\boldsymbol{\times}$') and the FBP (green, `{\color{green}$\blacksquare$}') reconstructions from real data of FAR and LAR scans specified in Table. \ref{table:LAR-views-real}. Solid red line indicates the corresponding reference value.}
\label{fig:ss-parameters-real}
\end{figure}

In Fig. \ref{fig:ss-parameters-real}, we display 
$\bar{\tau}_k$s and $\bar{\eta}_k$s, along with their standard deviations, estimated from DTV and FBP reconstructions of the SS images from FAR and LAR data. It can be observed that $\bar{\tau}_k$s and $\bar{\eta}_k$s from the DTV reconstructions for LAR scans agree reasonably well with the corresponding $\tau^{(\rm e)}_k$ and $\eta^{(\rm e)}_k$. Conversely, $\bar{\tau}_k$s and $\bar{\eta}_k$s obtained from the FBP reconstructions appear to have significant biases
 relative to $\tau^{(\rm e)}_k$ and $\eta^{(\rm e)}_k$ for LAR scans, especially for scans LAR3-LAR8.

\section{Discussion}\label{sec:discussion}
In the work, we investigate and develop the DTV algorithm for reconstruction of SS images directly from LAR data in CW-ZM EPRI. The work is motivated by (1)  the observation that, because the physical gradient strength of the spatial encoding magnetic field applied can be only of finite value, data cannot be acquired physically in the vicinity of $\vert \gamma\vert = 90^\circ$, thus resulting in an inherent LAR reconstruction problem; 
(2)  the interest in reconstruction of an SS image directly from data in CW-ZM EPRI; and (3) the interest in the development of CW-ZM EPRI with data acquired only over LARs for possibly minimizing the scan time. Based upon the DD-data model in Eq. \eqref{eq:dd-model-A} for CW-ZM EPRI, we first formulate the reconstruction problem as a constrained optimization program with unique constraints on the DTVs of an SS image and then developed the DTV algorithm to reconstruct the SS image through solving the optimization program. 

We carry out simulated- and real-data studies to verify and evaluate, in terms of visual inspection and quantitative analysis, the DTV algorithm in reconstructions of SS images directly from data acquired in LAR scans of varying angular ranges in CW-ZM EPRI.
The study results reveal that the DTV algorithm can accurately reconstruct SS images directly from data collected in LAR scans of angular ranges considerably smaller than that of the FAR scan,  that it can stably reconstruct SS images directly from real LAR data collected in CW-ZM EPRI, and that it can yield, from FAR data, SS images of quality improved over those  obtained with the FBP algorithm used in current CW-ZM EPRI. The DTV constraints are demonstrated to be effective in suppression of LAR and other artifacts that are observed otherwise in images reconstructed with the FBP algorithm.

While the spectral shape described in Eq. \eqref{eq:voigt} is considered in the studies, we note that the optimization-based reconstruction approach and the associated DTV algorithm proposed can readily accommodate spectral shapes of other forms \cite{sotgiu1987esr,kuppusamy1995mapping,ardenkjaer1998epr}, including regular and distorted/broadened Lorentzians. In the simulated-data study, we have chosen the DTVs of the truth SS image, i.e., the digital phantom, computed as the values of the DTV-constraint parameters, whereas in the real-data study, we have selected the constraint parameter values that yield the SS images with visually minimal artifacts. However, in an application with a specific goodness metric devised, 
the values of the constraint parameters can be chosen empirically for maximizing the metric. In particular, one may perform reconstructions with multiple sets of the constraint-parameter values chosen, then compute the metric values from the reconstructions, and finally select the set of constraint-parameter values yielding the numerically highest value of the application metric. We also note that the reconstruction approach and DTV algorithm developed can readily be applied to sampling schemes other than the ESA-sampling scheme considered.  

We have performed additional quantitative analysis with digital and physical phantoms of anatomies different than those in Figs. \ref{fig:image} and \ref{fig:phy-phan} in simulated- and real-data studies. Based upon the results obtained, observations can be made similar to those discussed in Sec. \ref{sec:results}. While the constrained optimization program designed in Eq. \eqref{eq:optA} is demonstrated to reduce LAR artifacts in SS images reconstructed from LAR data, it remains of interest to investigate different designs of optimization programs for potentially further minimizing artifacts for SS images reconstructed from data of LAR scans with angular ranges even smaller than those considered. For example, the data-fidelity term considered is of the $\ell_2$-norm form in the work,
it is worthy to investigate data-fidelity terms of additional forms \cite{zhang2016artifact} such as the Kullback-Leibler-divergence form for possibly reducing artifacts in SS images reconstructed from data collected in LAR scans of very small angular ranges.

It is worthy to note that the study design and DTV algorithm developed in the work can naturally be extended to investigating SS image reconstruction from FAR or LAR data acquired at sparse views \cite{Sidky:08, christodoulou2016fast,johnson2014compressed,qiao2018optimization} in CW-ZM EPRI. It remains a future task of interest to tailor, optimize, and evaluate image reconstructions from sparse-view data by using the DTV algorithm developed in CW-ZM EPRI.

\section{Conclusion}\label{sec:conclusion}
We have investigated and developed the optimization-based DTV algorithm for accurate reconstruction of SS images directly from  FAR and LAR data in CW-ZM EPRI. Results of simulated- and real-data studies reveal that from LAR data, SS images may be reconstructed that are visually and quantitatively comparable to those obtained from FAR data in CW-ZM EPRI. The DTV algorithm may be exploited possibly for enabling CW-ZM EPRI with reduced imaging time and artifacts by acquiring only LAR data instead of FAR data. The approach and DTV algorithm can readily be tailored to reconstruct SS images directly from FAR and LAR data in CW EPRI with scanning schemes other than the ZM scheme.

\section*{Declaration of Competing Interest}
Authors HH and BE declare US patents 8,664,955, 10,568,537, and 9,392,957 on aspects of the pO2 imaging technology and memebership in a start up company O2MThe o to market the pO2 imaging technology. Other authors declare that they have no known competing financial interests or personal relationships that could have appeared
to influence the work reported in this paper. 

{\color{black}
\section*{Data availability}
The authors will provide the reconstruction code used in this study upon request.
}

\section*{Acknowledgments}
This work was supported in part by NIH Grants R01EB026282, R01EB023968, R21CA263660, P41EB02034, R01CA98575, and R01CA236385. The computation of the work was performed in part on Computer Cluster Mel in the Department of Radiology at The University of Chicago. The contents of this article are solely the responsibility of the authors and do not necessarily represent the official NIH views. {\color{black} We would like to thank the anonymous reviewers for their insightful and constructive comments.}

\appendix
\addcontentsline{toc}{section}{Appendices}

\section{Continuous-to-continuous (CC)-data model in CW-ZM EPRI}\label{sec:cc-model}
We summarize the data model in a continuous form in CW-ZM EPRI, which allows for an intuitive incorporation of underlying physics involved in CW-ZM EPRI. While a data model in a continuous form is not directly applicable to data acquired and an SS image reconstructed on discrete arrays in practical EPRI, it provides a basis for devising a DD-data model that can adequately relate data and an SS image on discrete arrays in practical CW-ZM EPRI. 

We consider an SS image of continuous form in a 4D-image space $\{x', y', z', B \}$ formed by 3D-spatial dimensions $x'$, $y'$, and $z'$ and 1D-spectral coordinate $B$. Assuming $L'$ mm, and $B_w$ G, to be the support along each of the 3D-spatial dimensions, and along the spectral dimension, respectively, within the 4D-image space, we form a 4D-physical support of size $L'\times  L'\times  L' \times B_w$ mm$^3$ G, with its center coinciding with the origin of the 4D-image space, and assume that the SS image of continuous form is completely enclosed in the support. It is noted that the units of the  3D-spatial and 1D-spectral dimensions are different. For discussion convenience, we introduce a new 4D space as  $\{x, y, z, B\}=\{c x', c y', c z', B\}$, where $c=B_w / L'$ G mm$^{-1}$, in which all of the dimensions are of identical unit G. Subsequently, we have a 4D-image support of size $L\times  L\times  L \times B_w$ G$^4$, where $L=c L'$, which completely encloses the SS image of continuous form. For example, assuming a spatial support of $2\times 2\times 2$ mm$^3$ and spectral support of 2 G, we have $c=1$ G/mm, and if $x'=0.41$ mm, $x=1 \,{\rm G/mm} \times 0.41 \,{\rm mm} = 0.41$ G. However, for clarity, we remain to use unit ``mm'' for each of spatial dimensions $x$, $y$, and $z$, but with the understanding that it is converted automatically to unit ``G'' in the mathematical analysis and algorithm development.

We also consider 4D-data space $\{\xi, \vec{\alpha}\}$, formed by scalar variable $\xi$ and 4D unit-vector $\hat{\alpha}=({\rm cos}\phi\,{\rm sin}\theta\,{\rm sin}\gamma,\, {\rm sin}\phi\,{\rm sin}\theta\,{\rm sin}\gamma,\, {\rm cos}\theta\,{\rm sin}\gamma,\, {\rm cos}\gamma)^\top$, where $\gamma$ depicts the angle between the $B$-axis and vector $\hat{\alpha}$ and takes a positive or negative value when the inner product of  $\hat{\rm e}_z$ and $\hat{\alpha}$ is positive or negative. We also use $\hat{\alpha}_p$ to denote the projection of $\hat{\alpha}$ onto the hyperplane specified by $B=0$ in the 4D-image space. Moreover, in 3D spatial space $\{x, y, z\}$ with unit vectors $\hat{\rm e}_x$, $\hat{\rm e}_y$, and $\hat{\rm e}_z$, we use $\theta$ to depict the angle between $z$-axis and $\hat{\alpha}_p$, which takes a positive or negative value when the inner product of $\hat{\rm e}_x$ and $\hat{\alpha}_p$ is positive or negative, and also $\phi$ to depict the angle between $x$-axis and the projection of $\hat{\alpha}_p$ onto the $x$-$y$ plane, which takes a positive or negative value when the inner product of $\hat{\rm e}_y$ and $\hat{\alpha}_p$ is positive or negative. Considering a 3D hyperplane in the 4D-image space, we use scalar variable $\xi$, and unit vector $\hat{\alpha}$, to specify its distance to the origin of, and its orientation in, the 4D image space. 

We use $g(\xi,\hat{\alpha})$ to denote the data function in CW-ZM EPRI, which can be expressed as \cite{williams2005epr}
\begin{eqnarray}\label{eq:cc-model}
g(\xi,\hat{\alpha})
& = & {\rm cos}\gamma \, \int d{\vec{r}}\, \frac{\partial f(\vec{r})}{\partial B} \,\delta(\xi-\vec{r}\cdot\hat{\alpha}) \nonumber\\
&=  &{\rm cos}\gamma \,\mathbb{R}\!\left[ \frac{\partial f(\vec{r})}{\partial B} \right], 
\end{eqnarray}
where $\vec{r}$ is a point in the 4D space, $\mathbb{R}$ denotes the 4D-Radon transform of $\frac{\partial f(\vec{r})}{\partial B}$, i.e., the partial derivative of 4D-SS image $f(\vec{r})$ of continuous form, over a hyperplane specified by $\xi$ and $\hat{\alpha}$, $\xi\in [-\xi_{\tau}, \xi_{\tau}]$ with $2 \xi_{\tau}\ge {\rm max}(L, B_w)$, and ${\rm max}(\cdot)$ returns the largest value.  In the FAR scan, $\gamma \in [-\frac{\pi}{2}, \frac{\pi}{2}]$, $\theta \in [-\frac{\pi}{2}, \frac{\pi}{2}]$, and  $\phi\in [-\frac{\pi}{2}, \frac{\pi}{2}]$. Because $\xi$, $\hat{\alpha}$, and $\vec{r}$ can continuously vary, the data model in Eq. \eqref{eq:cc-model} is thus referred to as the continuous-to-continuous (CC)-data model.

In CW-ZM EPRI, the magnetic field gradient is expressed as $\mathbf{G}=G\hat{\mathbf{G}}$, where $G$ represents the magnitude of the vector with units of G/mm, and unit vector $\hat{\mathbf{G}}$ is dimensionless specifying the direction of $\mathbf{G}$. Angle $\gamma$ is defined as $\vert\gamma\vert={\rm arctan}(G/c)$, while  angles $\theta$ and $\phi$ specify $\hat{\mathbf{G}}$.
Algorithms have been developed for obtaining 4D-PD-SS image $\frac{\partial f(\vec{r})}{\partial B}$ through inverting the 4D-Radon transform in Eq. \eqref{eq:cc-model} from its full knowledge in
$\gamma \in [-\frac{\pi}{2}, \frac{\pi}{2}]$, $\theta \in [-\frac{\pi}{2}, \frac{\pi}{2}]$, and  $\phi\in [-\frac{\pi}{2}, \frac{\pi}{2}]$.  In practical CW-ZM EPRI, the physically achievable maximum gradient strength $G_{\tau}$ must be finite, and thus knowledge of $g(\xi,\hat{\alpha})$ can be measured only over $\gamma\in [-\gamma_{\tau}, \gamma_{\tau}]$, where $\gamma_{\tau}={\rm arctan}(G_{\tau}/c) < 90^\circ$, instead of over full angular range $\gamma \in [-\frac{\pi}{2}, \frac{\pi}{2}]$. Therefore, image reconstruction in practical CW-ZM EPRI is inherently a reconstruction problem from LAR data.  In Sec. \ref{sec:dd-data-form}, we use the CC-data model in Eq. \eqref{eq:cc-model} to devise a DD-data model in Eq. \eqref{eq:dd-model-A} adopted in the work.

\section{Convergence of the DTV algorithm}\label{sec:DTV-convergence-conditions}
We devise convergence conditions for the DTV algorithm and verify that  the DTV algorithm can reconstruct accurately an SS image from its FAR data through solving the optimization program in Eq. \eqref{eq:optA}. 
 
\subsection{Convergence conditions of the DTV algorithm}\label{sec:DTV-necessary-convergence-conditions}
We devise below dimensionless convergence metrics as 
\begin{equation}\label{eq:convergence-SS}
\begin{split}
\widetilde{D}^{(n)}_\mathbf{g} &= |\sqrt{D_{\mathbf{g}}(\mathbf{f}^{(n)})}|/||\mathbf{g}^{[\mathcal{M}]}||_2\\
 \widetilde{D}^{(n)}_{\rm TV_{\it x}} &= | (||\mathcal{D}_x \mathbf{f}^{(n)}||_1 - t_x)|/t_x \\
 \widetilde{D}^{(n)}_{\rm TV_{\it y}} &= | (||\mathcal{D}_y \mathbf{f}^{(n)}||_1 - t_y)|/t_y \\
 \widetilde{D}^{(n)}_{\rm TV_{\it z}} &= | (||\mathcal{D}_z \mathbf{f}^{(n)}||_1 - t_z)|/t_z \\
 \widetilde{D}^{(n)}_{\rm TV_{\it B}} &= | (||\mathcal{D}_B \mathbf{f}^{(n)}||_1 - t_B)|/t_B, \\
\end{split}
\end{equation}
where ${D}_\mathbf{g}({\mathbf{f}^{(n)}})=\frac{1}{2} \!\parallel \mathcal{C}\mathcal{R}\mathcal{D}_B {\mathbf{f}^{(n)}} - \mathbf{g}^{[\mathcal{M}]} \parallel_2^2 $,  $\mathbf{f}^{(n)}$ denotes the SS image reconstructed at iteration $n$, and $\mathbf{g}^{[\mathcal{M}]}$ depicts measured data. In the simulated-data study considered in the work,  $\mathbf{g}^{[\mathcal{M}]}=\mathbf{g}$ obtained with the truth SS image in Eq. \eqref{eq:dd-model-A}, whereas in the real-data study, $\mathbf{g}^{[\mathcal{M}]}$ are acquired in the experiment, which contain physical factors such as noise that are not included in (and thus are inconsistent with) the DD-data model in Eq. \eqref{eq:dd-model-A}.
The necessary convergence conditions for the DTV algorithm thus are given by 
\begin{equation}\label{eq:convergence_1}
\widetilde{D}^{(n)}_\mathbf{g}\rightarrow C,  \,\,\,\,\,\,  \widetilde{D}^{(n)}_{\rm TV_{\it x}}\rightarrow 0,  \,\,\,\,\,\,  \widetilde{D}^{(n)}_{\rm TV_{\it y}}\rightarrow 0,\,\,\,\,\,\,  \widetilde{D}^{(n)}_{\rm TV_{\it z}}\rightarrow 0,
\,\,\,\,\,\, \widetilde{D}^{(n)}_{\rm TV_{\it B}}\rightarrow 0,
\end{equation}
as $n\rightarrow \infty$, where $C$ is a constant that is 0 for simulated data or $>0$ for real data, respectively.

\renewcommand{\thefigure}{B.\arabic{figure}}
\setcounter{figure}{0}
\begin{figure}
\centering
\includegraphics[angle=0,trim=0 0 0 0, clip,origin=c,width=0.8\textwidth]{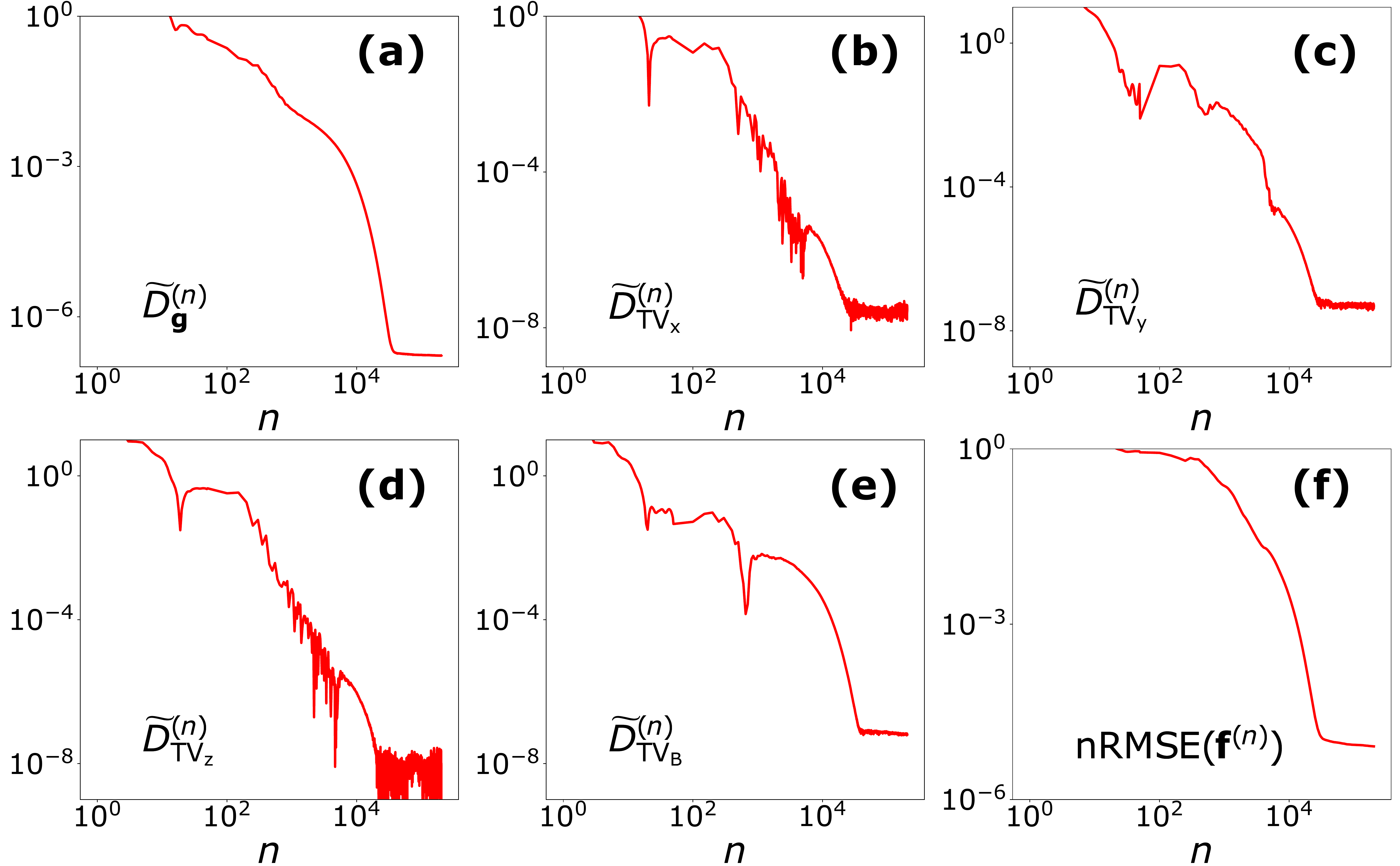}
\caption{Convergence metrics (a) $\widetilde{D}^{(n)}_\mathbf{g}$, (b) $\widetilde{D}^{(n)}_{\rm TV_{\it x}}$, (c) $\widetilde{D}^{(n)}_{\rm TV_{\it y}}$, (d) $\widetilde{D}^{(n)}_{\rm TV_{\it z}}$, (e) $\widetilde{D}^{(n)}_{\rm TV_{\it B}}$, and (f) ${\rm nRMSE}(\mathbf{f}^{(n)})$
of the DTV algorithm, as functions of iteration number $n$ (in log-log scales).}
\label{fig:invcrm_f}
\end{figure}

\subsection{Numerical verification of the DTV algorithm}\label{sec:verification-study}

It is critical to verify that the DTV algorithm and its computer implementation can accurately solve the optimization program in Eq. \eqref{eq:optA} in terms of numerically achieving the convergence conditions in Eq. \eqref{eq:convergence_1}. We have conducted extensive verification studies by using FAR and LAR data in both simulated- and real-data studies. Without loss of generality, we summarize here the verification study involving FAR data in the simulated-data study.

In Figs. \ref{fig:invcrm_f}a-\ref{fig:invcrm_f}e, the convergence metrics of the DTV algorithm are shown as functions of iteration number $n$ (in log-log scales); and these convergence curves clearly demonstrate that the DTV algorithm (and its computer implementation) can accurately solve the optimization program in Eq. \eqref{eq:optA} by achieving the necessary convergence conditions in Eq. \eqref{eq:convergence_1}. As mentioned above, convergence results similar to those in Fig. \ref{fig:invcrm_f} are obtained also for LAR scans in the simulated-data study and for FAR and LAR scans in the real-data study.

Furthermore, we also plot in Fig. \ref{fig:invcrm_f}f metric ${\rm nRMSE}(\mathbf{f}^{(n)})$ of Eq. \eqref{eq:imdist} in which the truth SS image is used as the reference image.  It can be observed that  the ${\rm nRMSE}(\mathbf{f}^{(n)})$ curve approaches zero as $n\rightarrow \infty$, suggesting that the final SS image reconstructed is numerically identical to the truth SS image. 
The result reveals empirically that the DTV algorithm (and its computer implementation) can reconstruct accurately an SS image from FAR data consistent with the DD-data model in Eq. \eqref{eq:dd-model-A}. In other words, the DTV algorithm and its computer implementation are verified empirically to invert system matrix $\mathcal{H}$ in the DD-data model in Eq. \eqref{eq:dd-model-A} for the FAR scan in CW-ZM EPRI. Once verified, the DTV algorithm (and its computer implementation) can be used for empirically investigating reconstructions of SS images from LAR data and/or from FAR and LAR data containing physical components such as noise inconsistent with the DD-data model based upon which the DTV algorithm is developed.


\end{document}